\documentclass[twocolumn,showpacs,preprintnumbers,amsmath,amssymb]{revtex4}
\usepackage{amsmath}
\usepackage{natbib}
\usepackage{graphicx}% Include figure files
\usepackage{epstopdf}
\usepackage{subfigure}
\usepackage{dcolumn}% Align table columns on decimal point
\usepackage{bm}% bold math
\usepackage{color}

\begin{document}

%\preprint{}

\title {Extended phase-space symplectic-like integrators
for coherent post-Newtonian Euler-Lagrange equations} % Force line breaks with \\
\author{Guifan Pan$^{1}$}%\email{1612110208@st.gxu.edu.cn}303131599@qq.com
\author{Xin Wu$^{1,2,3}$}
\email{wuxin_1134@sina.com}
\author{Enwei Liang$^{1,3}$}
%\email{lew@gxu.edu.cn}

\affiliation{1. School of Physical Science and Technology, Guangxi
University, Nanning 530004, China \\ 2. School of Mathematics,
Physics and Statistics $\&$ Center of Application and Research of
Computational Physics, Shanghai University of Engineering Science,
Shanghai 201620, China \\ 3. Guangxi Key Laboratory for
Relativistic Astrophysics, Guangxi University, Nanning 530004,
China}

%%\date{\today} % It is always \today, today,

\begin{abstract}

Coherent or exact equations of motion for a post-Newtonian
Lagrangian formalism are the Euler-Lagrange equations without any
terms truncated. They naturally conserve  energy {and} angular
momentum. Doubling the phase-space variables of positions and
momenta in the coherent equations, we establish extended
phase-space symplectic-like integrators with the midpoint
permutations. The velocities should be solved iteratively from the
algebraic equations of the momenta defined by the Lagrangian
during the course of numerical integrations. It is shown
numerically that a fourth-order extended phase-space
symplectic-like method exhibits good long-term stable error
behavior in energy {and} angular momentum, as a fourth-order implicit
symplectic method with a symmetric composition of three
second-order implicit midpoint rules or a fourth-order
Gauss-Runge-Kutta implicit symplectic scheme does. For given time
step and integration time, the former method is superior to the
latter integrators in computational efficiency.
{The extended phase-space method is used to study
the effects of the parameters and initial conditions on the
orbital dynamics of the coherent Euler-Lagrange equations for a
post-Newtonian circular restricted three-body problem. It is also
applied to trace the effects of the initial spin angles, initial
separation and initial orbital eccentricity on the dynamics of the
coherent post-Newtonian Euler-Lagrange equations of spinning
compact binaries. }

%\textbf{Keywords}: Lagrangian  formalism, Euler-Lagrange
%equations, post-Newtonian approximations, chaos

\end{abstract}

%%\pacs{04.25.Nx, 04.25.-g, 45.20.Jj}% PACS, the Physics and Astronomy
% Nonlinear dynamics and chaos 05.45.-a,
% 45.20.Jj Lagrangian and Hamiltonian mechanics  ,
%95.10.Fh Chaotic dynamics

\maketitle%
\section{Introduction}

Gravitational waves and black holes are two fundamental
predictions of the theory of Einstein's general relativity. The
predictions were recently confirmed by a series of detections of
gravitational waves, such as GW150914 \cite{1} and GW190521 [2,
3]. {Chaos is a possibly terrible obstacle to a
method of matched filtering, which requires a gravitational wave
signal drawn out of the noise in excellent agreement with a
theoretical template of the gravitational wave [4]. Thus, chaos in
systems of spinning compact binaries has been studied by several
authors [4-14]. The chaotic behavior in these references was
mainly considered in conservative binary systems. In fact, chaos
may be irrelevant in dissipative binary black hole systems due to
the fast dissipation [15]. }

Higher-order post-Newtonian (PN) approximations are well
applicable to the description of relativistic two-body dynamics of
binary black hole mergers and to accurate predictions of the
gravitational waveforms from the binary mergers. There are two
different PN approximation methods. One method is the
Arnowitt-Deser-Misner (ADM)-Hamiltonian formalism of general
relativity for describing the motion of two compact bodies in the
ADM coordinates [16-18]. The other method is the equations of
motion in harmonic coordinates and the Lagrangian of the motion
that is deduced from the equations of motion [19-23]. The physical
equivalence of the ADM-Hamiltonian formalism and the
harmonic-coordinates Lagrangian formalism at same PN order was
shown by several authors [23-25]. However, the authors of [26]
{claimed that the two PN formalisms are not
exactly} equivalent in the orbital dynamical behavior. In general,
the higher-order PN terms are truncated when one of the two
formalisms is transformed into the other one through the Legendre
transformation. These truncated terms are regarded as a difference
between the two PN approximation formalisms [26]. This difference
is negligible for a weak gravitational field like the solar
system, but may exert an important influence on the dynamics of
the two formalisms for a strong gravitational field of compact
objects. In some cases, the two formalisms in same coordinate
system under same coordinate gauge have different orbital
dynamical behaviors. {The authors of [27, 28] showed the
integrability and non-chaoticity of the conservative 2PN
ADM-Hamiltonian dynamics of compact binaries with leading order
spin-orbit interaction when the binaries are spinning. However,
chaos exists in the PN Lagrangian systems of compact binaries
having two arbitrary spins with spin-orbit interactions [14]. In
addition, the 1PN Lagrangian and Hamiltonian dynamics for a
circular restricted three-body problem of compact objects may be
different from qualitative and quantitative perspectives [26].}

{There are} two paths for deriving the equations
of motion (i.e., the Euler-Lagrange equations) from a conservative
PN Lagrangian formalism. One path gives the equations of motion
remaining at the same PN order of the Lagrangian by truncating the
higher-order PN terms in the Euler-Lagrange equations. In fact,
the equations of motion are approximately derived are called as
approximate PN Lagrangian equations of motion. The other path is
the differential equations of positions and generalized momenta,
where velocities are solved from the algebraic equations of the
generalized momenta via an iterative method. Such equations of
motion are coherent (or exact) PN Lagrangian equations of motion
[29, 30]. The approximate Lagrangian equations approximately
conserve the integrals of motion (e.g., energy) in the PN
Lagrangian formalism in general, but the exact ones always
strictly conserve the integrals of motion from a theoretical point
of view. {Dubeibe et al. [31] investigated the
conservation of the Jacobi integral in a PN circular restricted
three-body problem, and found that the approximate Euler-Lagrange
equations do not not conserve the Jacobi integral for the distance
$a$ of the two primaries and the speed of light $c$ satisfying the
conditions $a=c=1$, but conserve well for the case of $a=1$ and
$c=10^{4}$. The authors of [29, 30] pointed out that the
approximate Lagrangian equations and the exact ones do not have
typical differences for the case of $a=1$ and $c=10^{4}$, but have
for the case of $a=c=1$. They numerically showed that the exact
Euler-Lagrange equations always well conserve the Jacobi integral
regardless of the choice of $a$ and $c$. The Hamiltonian formalism
can also strictly conserve the integrals of motion from the
theory. Dubeibe et al. [31] numerically confirmed that the
conservation of the Jacobi integral in the Hamiltonian is
independent of the choice of $a$ and $c$. In a word, the
approximate Euler-Lagrange equations, exact Euler-Lagrange
equations and Hamiltonian equations at same PN orders have
explicit differences for a strong gravitational field of compact
objects, whereas have negligible differences for a weak
gravitational field like the solar system [26, 29, 30, 32]. The
approximate Euler-Lagrange equations for a given PN Lagrangian
formalism of compact objects are a set of wrong equations of
motion and poorly conserve the constants of motion. Instead, the
exact Euler-Lagrange equations should be used. }

When the  binaries without spin, the conservative Lagrange
formalism with the exact Euler-Lagrange equations and the
conservative Hamiltonian formalism up to any PN order have four
integrals of motion including the total energy and the total
angular momenta. Hence the orbits of it are integrable and
regular. The presence of chaos in [26, 14, 4, 5, 7-13]
{arises} due to the spins of the binary destroying
the integrability of PN systems of compact binaries. The
canonical, conjugate spin variables of Wu and Xie [33] play an
important role in determining the integrability or
nonintegrability of Hamiltonian systems of spinning compact
binaries. When only one of the binary objects spins, any
conservative PN Hamiltonian with 8 dimensions is always integrable
and nonchaotic regardless of PN orders and spin effects due to the
total energy and the total angular momenta as four constants of
motion. {Because a conservative PN Lagrangian
approach of compact binaries with one body spinning can be
formally equivalent to an integrable PN Hamiltonian, it is
impossibly chaotic.} As an important result of [32], no chaos
occurs in any conservative PN Lagrangian and Hamiltonian
approaches when only one body of comparable mass binaries spins.
This result significantly improve on clarifying the doubt on the
presence or absence of chaos in conservative PN Lagrangian and
Hamiltonian approaches of compact binaries with one body spinning
in [27, 28, 14, 27]. In addition, conservative PN Hamiltonian
systems of compact binaries having two arbitrary spins with
spin-orbit interactions were given parametric solutions in [27,
28]. Because they hold five integrals of motion consisting of the
total energy, the total angular momenta and the magnitude of the
Newtonian-like angular momenta in 10-dimensional phase space, and
are integrable. The PN Lagrangian systems of compact binaries
having two arbitrary spins with spin-orbit interactions lead to
the 3PN spin-spin couplings in the equivalent Hamiltonian
formalisms (note: the PN order of the Hamiltonians is unlike that
of the Lagrangians). The spin-spin couplings do not conserve the
magnitude of the Newtonian-like angular momenta, therefore, the
Hamiltonian formalisms (i.e., Lagrangian systems with  the exact
Euler-Lagrange equations) are nonintegrable and probably chaotic
[26]. This is why the spin-orbit interactions can produce chaos in
the Lagrangians but cannot in the Hamiltonians.

Usually, a long enough time integration is necessary to detect the
chaotical behavior. Such a numerical integration scheme should
have high precision, good stability and small expense of
computational time. Low order geometric integrators [34] can
provide reliable results and take less computational cost in the
case of long-term integration, and therefore they are naturally
chosen. When dealing with Hamiltonian systems, the most
appropriate geometric integrators are symplectic schemes, which
preserve the symplectic nature of Hamiltonian dynamics and have no
secular drift in energy errors. Due to inseparable variables,
completely implicit symplectic methods, such as
the implicit midpoint method [35] and the Gauss-Legendre
Runge-Kutta implicit symplectic methods [36] or Gauss-Runge-Kutta
(GRK) implicit symplectic methods [34,37-39], are mainly used in
relativistic spacetimes or PN systems. There are also explicit and
implicit combined symplectic methods [40-43]. More recently,
explicit symplectic integrators were successfully designed for the
dynamics of charged particles around the Schwarzschild,
Reissner-Nordstr\"{o}m, Reissner-Nordstr\"{o}m-(anti)-de Sitter
and Kerr black holes, and these  black holes  surrounded with
external magnetic fields [40-47]. However, these explicit
symplectic integrators are difficultly applied to PN systems of
spinning compact binaries. A notable point is that the
computational cost is generally less for explicit symplectic
integrators than for implicit ones. When the phase space of an
inseparable Hamiltonian system is extended, standard explicit
symplectic leapfrog splitting methods are still available [48].
However, the extended phase-space leapfrogs are not symplectic due
to phase space mixing and projection. In spite of this, the
extended phase space leapfrogs have symmetries and then show good
long-term stability and error behavior. In this sense, the
algorithms are called as symplectic-like schemes. Optimal choices
of mixing maps were considered in Refs. [49, 50]. The extended
phase space symplectic-like schemes with optimal mixing maps were
applied to PN systems of spinning compact binaries and other
inseparable Hamiltonian problems [51-54].

The main aim of the present paper is to discuss a possible
application of extended phase space symplectic-like integrators to
the coherent PN Euler-Lagrange  equations [29, 30]. For this
purpose, we construct symplectic-like integrators for the exact PN
Euler-Lagrange equations in an extended phase space of a PN
Lagrangian formalism in Sect. 2. Then, one of the extended phase
space symplectic-like integrators is applied to the exact PN
Euler-Lagrange equations of a PN circular restricted three-body
problem [55], and the dynamics of the problem is explored in Sect.
3. It is further applied to the exact PN Euler-Lagrange equations
of spinning compact binaries [56], and the dynamics of spinning
compact binaries is investigated in Sect. 4. Finally, the main
results are concluded in Sect. 5.

\section{Extended phase-space symplectic-like
integrators}

The coherent equations of motion for a PN Lagrangian system are
introduced. Then, symplectic-like integrators in extended phase
space of the coherent equations are constructed.

\subsection{Coherent PN Euler-Lagrange equations}

Suppose $\mathcal{L}(\mathbf{r},\mathbf{v})$ is a Lagrangian
formulation of PN order $j$, where $\mathbf{r}$ and $\mathbf{v}$
denote position and velocity vectors, respectively. Based on this
Lagrangian, a  generalized momentum vector is defined as
\begin{equation}
\mathbf{p}=\frac{\partial\mathcal{L}}{\partial\mathbf{v}}=\mathbf{P}(\mathbf{r},\mathbf{v}).
\end{equation}
It is easy to exactly express $\mathbf{p}$ in terms of
$\mathbf{r}$ and $\mathbf{v}$. Inversely, it may not be easy to
write an exact expression of $\mathbf{v}$ in terms of $\mathbf{r}$
and $\mathbf{p}$. In classical mechanics, $\mathbf{v}$ can be
described exactly in general. However, the exact description  of
$\mathbf{v}$ often becomes difficult in general relativity or
relativistic PN approximations  because $\mathbf{p}$ is a
nonlinear function of $\mathbf{v}$ in most cases. There are two
paths for the description of $\mathbf{v}$. Path 1 is based on the
PN approximations and obtains
\begin{equation}
\mathbf{v}=\mathbf{V}(\mathbf{r},\mathbf{p})+\mathcal{O}(\frac{1}{c^2})^{j+1},
\end{equation}
where $c$ is the speed of light. Path 2 is iteratively solving a
modified version of  the algebraic equation (1):
\begin{equation}
\mathbf{v}=\mathbf{Q}(\mathbf{r},\mathbf{v},\mathbf{p}).
\end{equation}
Here, the Newton or Seidel iteration method is adopted. The PN
Lagrangian corresponds to energy
\begin{equation}
E(\mathbf{r},\mathbf{v})=\mathbf{v}\cdot
\mathbf{p}-\mathcal{L}(\mathbf{r},\mathbf{v}).
\end{equation}

If the velocity from the PN approximation in Eq. (2) is
substituted into Eq. (4), then there is a Hamiltonian of PN order
$j$
\begin{equation}
H(\mathbf{r},\mathbf{p})=\mathbf{v}\cdot
\mathbf{p}-\mathcal{L}(\mathbf{r},\mathbf{v}).
\end{equation}
{Eq. (5) is a standard Hamiltonian, which is a
function of the coordinates and momenta. The Hamiltonians
mentioned in the Introduction are such standard Hamiltonians.}
Because the PN terms higher than the $j$th order are truncated in
Eq. (2), $H$ is not exactly but is approximately equal to $E$:
\begin{equation}
H(\mathbf{r},\mathbf{p})= E(\mathbf{r},\mathbf{v})
+\mathcal{O}(\frac{1}{c^2})^{j+1}.
\end{equation}

When the velocity obtained from an iterative solution of Eq. (3)
is substituted into Eq. (4), there is still a $j$th order PN
Hamiltonian
\begin{equation}
\mathcal{H}(\mathbf{r},\mathbf{v},\mathbf{p})=\mathbf{v}\cdot
\mathbf{p}-\mathcal{L}(\mathbf{r},\mathbf{v}),
\end{equation}
where $\mathbf{v}$ is an implicit function of $\mathbf{r}$ and
$\mathbf{p}$. {Eq. (7) is a \emph{formal}
Hamiltonian, which is a function of the coordinates, momenta and
velocities.} Note that Eqs. (4), (5) and (7) are the Legendre
transformation.

Clearly, the difference between Eqs. (5) and (7) is that
$\mathbf{v}$  is expressed in terms of $\mathbf{p}$ in Eq. (5),
but it is not in Eq. (7). In fact, Eq. (7) is equivalent to Eq.
(4) in the expressional form. Thus, $\mathcal{H}$, $H$ and $E$
satisfy the relations
\begin{equation}
\mathcal{H}=E=H +\mathcal{O}(\frac{1}{c^2})^{j+1}.
\end{equation}
That is, $\mathcal{H}=E\approx H$.

The canonical equations for the Hamiltonian $H$ in Eq. (5) are
written as follows:
\begin{eqnarray}\label{Eq:9}
\dot{\mathbf{r}} &=& \frac{\partial H }{\partial \mathbf{p}} \\
\dot{\mathbf{p}} &=& -\frac{\partial H }{\partial \mathbf{r}}.
\end{eqnarray}
They exactly conserve the Hamiltonian $H$ rather than the energy
$E$ or  the Hamiltonian $\mathcal{H}$. The Hamiltonian
$\mathcal{H}$ corresponds to the canonical equations
\begin{eqnarray}
\dot{\mathbf{r}} &=& \frac{\partial\mathcal{H} }{\partial \mathbf{p}}=\mathbf{v} \\
\dot{\mathbf{p}} &=& -\frac{\partial\mathcal{H} }{\partial
\mathbf{r}}=\mathbf{P}(\mathbf{r},\mathbf{v}).
\end{eqnarray}
Eq. (12) is the Euler-Lagrange equation. The equation and Eqs. (3)
and (11) were called as the coherent or exact PN Lagrangian (or
Euler-Lagrange) equations of motion in [29, 30]. Naturally, the
coherent equations have the consistency of energy $E$ or
Hamiltonian $\mathcal{H}$. If $\mathbf{p}$ in Eq. (1) is
substituted into Eq. (12), the acceleration to the order $j$ reads
\begin{eqnarray}
\dot{\mathbf{v}}
=\mathbf{a}(\mathbf{r},\mathbf{v})+\mathbf{P}(\mathbf{r},\mathbf{v})+\mathcal{O}(\frac{1}{c^2})^{j+1},
\end{eqnarray}
where the acceleration $\mathbf{a}(\mathbf{r},\mathbf{v})$ is due
to the contribution of the derivative of momenta with respect to
time, $d\mathbf{p}/dt$, and remains at the order $j$. None of
$\mathcal{H}$, $H$ and $E$ can be conserved exactly by Eq. (13)
with Eq. (11). In fact, no one knows what energy is conserved
exactly by Eqs. (11) and (13).  Eqs. (11) and (13) are called as
the approximate Euler-Lagrange equations. The angular momentum can
not be maintained by the approximate equations, either.

The above demonstrations clearly show that the three sets of
motion equations, Eqs. (9) and (10), Eqs. (3), (11) and (12), and
Eqs. (11) and (13), have some differences although they are
accurate to the PN order $j$. They also exist slight differences
in the conservation of the integrals of motion. In particular, the
coherent Eqs. (3), (11) and (12) and the approximate Eqs. (11) and
(13) are different in computations. $\mathbf{v}$ is constant, and
$\mathbf{r}$ and $\mathbf{p}$ are integration variables during a
step integration of  the coherent equations. Once the solutions of
$\mathbf{r}$ and $\mathbf{p}$ are obtained after this step,
$\mathbf{v}$ is given by  solving the iterative equation (3).
However, $\mathbf{v}$ with $\mathbf{r}$ is directly integrated in
the approximate equations.

\subsection{Construction of symplectic-like integrators in extended
phase space}

Extending the phase space of a Hamiltonian, Pihajoki [48] proposed
explicit symplectic-like integrators for an extended phase-space
Hamiltonian. Following this idea, we consider an extension to the
phase space of coherent PN Lagrangian equations, i.e., the
Hamiltonian in Eq. (7). This extension is implemented by a new
Hamiltonian
\begin{eqnarray}
& & \Gamma\left(\mathbf{r},\mathbf{r}^{*},\mathbf{v},\mathbf{v}^{*},\mathbf{p},\mathbf{p}^{*}\right)  \nonumber \\
&& =\mathcal{H}_{A}\left(\mathbf{r},\mathbf{v}^{*},
\mathbf{p}^{*}\right) +\mathcal{H}_{B}
\left(\mathbf{r}^{*},\mathbf{v},\mathbf{p}\right),
\end{eqnarray}
where $\mathcal{H}_{A}$ and $\mathcal{H}_{B}$ are two
sub-Hamiltonians
\begin{eqnarray}
\mathcal{H}_{A} &=& \mathbf{p}^{*}\cdot\mathbf{v}^{*}-\mathcal{L}_{A}(\mathbf{r},\mathbf{v}^{*}), \\
\mathcal{H}_{B} &=&
\mathbf{p}\cdot\mathbf{v}-\mathcal{L}_{B}(\mathbf{r}^{*},\mathbf{v}).
\end{eqnarray}
Here, $\mathbf{v}^{*}$ is a function of $\mathbf{r}$ and
$\mathbf{p}^{*}$, and $\mathbf{v}$ is a function of
$\mathbf{r}^{*}$ and $\mathbf{p}$. They are still solved by the
iterative equations
\begin{eqnarray}
\mathbf{p}^{*} &=&\frac{\partial\mathcal{L}_{A}(\mathbf{r},\mathbf{v}^{*})}{\partial\mathbf{v}^{*}}
\rightarrow \mathbf{v}^{*}=\mathbf{Q}_{A}(\mathbf{r},\mathbf{v}^{*},\mathbf{p}^{*})\\
\mathbf{p}
&=&\frac{\partial\mathcal{L}_{B}(\mathbf{r}^{*},\mathbf{v})}{\partial\mathbf{v}}
\rightarrow
\mathbf{v}=\mathbf{Q}_{B}(\mathbf{r}^{*},\mathbf{v},\mathbf{p}).
\end{eqnarray}
$\mathcal{F}_{A}$ is an operator for iteratively solving Eq. (17),
and $\mathcal{F}_{B}$ is another operator for iteratively solving
Eq. (18). In fact, the two independent sub-Hamiltonians are the
same as the original Hamiltonian $\mathcal{H}$ in the expressional
forms. The two sub-Hamiltonians always satisfy the relation
$\mathcal{H}_{A}=\mathcal{H}_{B}$ for any time if their initial
conditions are the same.

$\mathcal{H}_{A}$ and $\mathcal{H}_{B}$ are independently
solvable. $e_{\mathcal{H}_{A}}^{h}$ is an operator for solving
$\mathcal{H}_{A}$, and  $e_{\mathcal{H}_{B}}^{h}$ is another
operator for solving $\mathcal{H}_{B}$, where $h$ represents a
step size. The solutions from $(n-1)$th step to $n$th step are
written as
\begin{equation}\label{Eq:5}
e_{\mathcal{H}_{A}}^{h}: \left(\begin{array}{l}
\mathbf{r}^{*}\\
\mathbf{p}\\
\end{array} \right)_{n}= \left(\begin{array}{l}
\mathbf{r}^{*} +h\cdot\mathbf{v}^{*} \\
\mathbf{p}+h\cdot\frac{\partial\mathcal{L}_{A}}{\partial\mathbf{r}}\\
\end{array}\right)_{n-1},
\end{equation}
\begin{equation}\label{Eq:6}
e_{\mathcal{H}_{B}}^{h}: \left(\begin{array}{l}
\mathbf{r}\\
\mathbf{p}^{*}\\
\end{array} \right)_{n}= \left(\begin{array}{l}
\mathbf{r}+h\cdot\mathbf{v}\\
\mathbf{p}^{*}+h\cdot\frac{\partial\mathcal{L}_{B}}{\partial\mathbf{r}^{*}}\\
\end{array}\right)_{n-1}.
\end{equation}

These operators symmetrically compose a second-order integration
algorithm
\begin{equation}\label{Eq:7}
A_{2}(h)= \left(\mathcal{F}_{A} \circ
e_{\mathcal{H}_{B}}^{h/2}\right)\circ \left(\mathcal{F}_{B}\circ
e_{\mathcal{H}_{A}}^{h}\right)\circ \left(\mathcal{F}_{A}\circ
e_{\mathcal{H}_{B}}^{h/2}\right).
\end{equation}
This construction is an explicit second-order symplectic  leapfrog
integrator if $\mathcal{F}_{A}$ and $\mathcal{F}_{B}$ are absent.
The second-order scheme can be used to yield a fourth-order method
of Yoshida [57]
\begin{equation}
A_{4}(h)=A_{2}(a_1 h)\circ  A_{2}(a_2h)\circ  A_{2}(a_1h),
\end{equation}
where $a_1=1/(2-2^{1/3})$ and $a_2=1-2a_1$. Algorithms $A_{2}$ and
$A_{4}$ are symplectic for the extended phase-space Hamiltonian
$\Gamma$. In these constructions, $\mathbf{r}^{*}$, $\mathbf{p}$,
$\mathbf{r}$ and $\mathbf{p}^{*}$ have explicit solutions, whereas
$\mathbf{v}$ and $\mathbf{v}^{*}$ must be solved iteratively.

A notable problem is that $\mathcal{H}_{A}$ and $\mathcal{H}_{B}$
have same solutions for same initial conditions when they are
independently solved, but have different solutions in these
algorithmic constructions because their solutions are coupled. To
avoid this problem as much as possible, Pihajoki [48] introduced
mixing maps (e.g., permutations of coordinates and/or momenta) as
a feedback between the two solutions. A  projection map on the
projection of a vector in extended phase space back to the
original phase space is also necessary. Liu et al. [52] showed
that sequent permutations of coordinates and momenta are a good
choice of the  mixing maps. Luo et al. [53] found that midpoint
permutations between coordinates and those between momenta are the
best choice of the mixing maps. The midpoint permutation map is
described by
\begin{equation}
\boldsymbol{M}=\left(\begin{array}{llll}
\frac{\mathbf{1}}{2}, & \frac{\mathbf{1}}{2}, & \mathbf{0}, & \mathbf{0} \\
\frac{\mathbf{1}}{2}, & \frac{\mathbf{1}}{2}, & \mathbf{0}, & \mathbf{0} \\
\mathbf{0}, & \mathbf{0}, & \frac{\mathbf{1}}{2}, & \frac{\mathbf{1}}{2} \\
\mathbf{0}, & \mathbf{0}, & \frac{\mathbf{1}}{2}, &
\frac{\mathbf{1}}{2}
\end{array}\right).
\end{equation}
When this map is included after algorithms $A_{2}$ and $A_{4}$, we
have new constructions. For example, $A_{4}$ becomes
\begin{equation}
EM4(h)= \boldsymbol{M}\circ  A_{4}(h).
\end{equation}
New method EM4 is still  accurate to the order of $h^4$.

The permutation map $\boldsymbol{M}$ is not symplectic, and then
EM4 is no longer symplectic. Even if $\boldsymbol{M}$ is
symplectic, EM4 is not for any choice of projection maps. However,
EM4 is time-symmetric, and therefore it may preserve the original
Hamiltonian without secular growth in the error, as a symplectic
scheme can. We check the numerical performance of EM4 using two PN
problems.

\section{PN circular restricted three-body problem}

A PN Lagrangian formulation of circular restricted three-body
problem is introduced in Sect. 3.1. The phase-space structures of
orbits in the  approximate Euler-Lagrange equations and the
coherent ones are described in Sect. 3.2. Then, energy errors for
four algorithms algorithms acting on the exact Euler-Lagrange
equations are compared. Sect. 3.3 relates to the dependence of
chaos on the initial value $x$ and the parameters in the coherent
Euler-Lagrange equations by use of the technique of fast Lyapunov
indicators (FLIs).

\subsection{Dynamical models}

Let us consider a PN  planar circular restricted three-body
problem.  Two primary bodies have separation $a$ and masses $M_1$
and $M_2$. Their total mass is $M=M_1+M_2$. The ratios of the two
bodies' masses to the total mass are $\mu_{i}=M_{i}/M (i=1,2)$.
The two primary bodies do circular motions. The angular speeds of
circular motions are defined as $\omega_{0}=\sqrt{GM/a^{3}}$ ($G$
being the gravitational constant) with respect to the barycenter
of the two bodies in the Newton gravity, and $\omega$ in the
relativistic gravity. The third body has a negligible mass $m$.
Its position and velocity $(x,y,v_x,v_y)$ in a rotating frame
evolve with time according to the dimensionless PN Lagrangian [55,
58]
\begin{equation}
\mathcal{L}=\mathcal{L}_{0}+ \mathcal{L}_{1}/c^{2}+
\mathcal{L}_{2}/c^{2},
\end{equation}
where the Newtonian Lagrangian system $\mathcal{L}_{0}$, and two
PN Lagrangian system $\mathcal{L}_{1}$ and $\mathcal{L}_{2}$ are
\begin{equation}\label{Eq:tb-L0}
\mathcal{L}_{0}=\frac{\mu_{1}}{d_{1}}+\frac{\mu_{2}}{d_{2}}+\frac{1}{2}\left(U^{2}+2
A \omega_{0}+R^{2} \omega_{0}^{2}\right),
\end{equation}
\begin{equation}\label{Eq:tb-L1}
\mathcal{L}_{1}=\omega_{0} \omega_{1}\left(A+R^{2}
\omega_{0}\right),
\end{equation}
\begin{equation}\label{Eq:tb-L2}\begin{aligned}
a\mathcal{L}_{2}=& \frac{1}{8}\left(U^{2}+2 A \omega+R^{2} \omega^{2}\right)^{2}-\mu_{1} \mu_{2}\left(\frac{1}{d_{1}}+\frac{1}{d_{2}}\right) \\
&-\frac{1}{2}\left(\frac{\mu_{1}}{d_{1}}+\frac{\mu_{2}}{d_{2}}\right)^{2}+\frac{3}{2}\left(\frac{\mu_{1}}{d_{1}}+\frac{\mu_{2}}{d_{2}}\right)\\
& \cdot\left(U^{2}+2 A \omega+R^{2} \omega^{2}\right)\\
&+\frac{3}{2}\omega^{2}\left(\frac{\mu_{1} x_{1}^{2}}{d_{1}}+\frac{\mu_{2} x_{2}^{2}}{d_{2}}\right)\\
&-\frac{7}{2} \omega(\dot{y}+\omega x)\left(\frac{\mu_{1} x_{1}}{d_{1}}+\frac{\mu_{2} x_{2}}{d_{2}}\right) \\
&-\frac{1}{2} \omega y(\dot{x}-\omega y) \cdot\left[\frac{\mu_{1} x_{1}\left(x-x_{1}\right)}{d_{1}^{3}} \right.\\
&\left.+\frac{\mu_{2} x_{2}\left(x-x_{2}\right)}{d_{2}^{3}}\right]\\
&-\frac{1}{2} \omega y^{2}(\dot{y}+\omega x)\left(\frac{\mu_{1}
x_{1}}{d_{1}^{3}}+\frac{\mu_{2} x_{2}}{d_{2}^{3}}\right).
\end{aligned}
\end{equation}
Here, $\omega_{1}=(\mu_{1}\mu_{2}-3)/(2a) $,
$\omega=\omega_{0}(1+\omega_{1}/c^{2})$,
$d_{1}=\sqrt{\left(x-x_{1}\right)^{2}+y^{2}}, \quad
d_{2}=\sqrt{\left(x-x_{2}\right)^{2}+y^{2}}$,
$R=\sqrt{x^{2}+y^{2}}, U^{2}=\dot{x}^{2}+\dot{y}^{2}, A=\dot{y}
x-\dot{x} y$, $x_1=-\mu_{2}$, and $x_2=\mu_{1}$. At the 1PN order,
the square of angular frequency is
$\omega^{2}=\omega_{0}^{2}(1+2\omega_{1}/c^{2})$. Geometric unit
$G=1$ is adopted. The above dimensionless operations are
implemented via a series of scale transformations [49]:
$\mathcal{L}\rightarrow m \mathcal{L}/a$, $t\rightarrow t
M/\omega_0$, $a\rightarrow M a$, $x\rightarrow Ma x$,
$x_1\rightarrow Ma x_1$, $x_2\rightarrow Ma x_2$, and
$y\rightarrow Ma y$. In this sense, $\dot{x}\rightarrow
a\omega_{0}\dot{x}$ and $\dot{y}\rightarrow a\omega_{0}\dot{y}$.

The coherent equations (3), (11) and (12) for $\mathcal{L}$ in Eq.
(25) can be easily written in detail.  The iteration equation (3)
in the present problem is written as
\begin{eqnarray}\label{Eq:iter}
v_{x} &=& \left(p_{x}-f_1(x,y,v_x,v_y)\right) \nonumber \\ &&
/\left[1+\frac{U^2}{2ac^2}+\frac{3}{ac^2}(\frac{\mu_1}{d_1}+\frac{\mu_2}{d_2}) \right], \\
v_{y} &=& \left(p_{y}-f_2(x,y,v_x,v_y)\right) \nonumber \\ &&
/\left[1+\frac{U^2}{2ac^2}+\frac{3}{ac^2}(\frac{\mu_1}{d_1}+\frac{\mu_2}{d_2})
\right],
\end{eqnarray}
where $f_1$ and $f_2$ are two known functions. The coherent
equations conserve the Jacobi constant
\begin{equation}\label{Eq:tb-cj}\begin{aligned}
C_{j}=-2E,
\end{aligned}\end{equation}
where $E$ is given by Eq. (4). On the other hand, the  approximate
PN Lagrangian equations of motion were given in [48, 51] by
\begin{equation}\label{Eq:tb-ap}\begin{aligned}
\dot{x}=&v_{x},\\
\dot{y}=&v_{y},\\
\dot{v}_{x}=&\left[2\dot{v_{y}}+x-\frac{\mu_{1}\left(x-x_{1}\right)}{d_{1}^{3}}
-\frac{\mu_{2}\left(x-x_{2}\right)}{d_{2}^{3}} \right]\\
&+2\omega_{1}\left(v_{y}+ x \right)/c^{2} + \mathcal{P}(x,y,v_x,v_y)/(ac^{2}),\\
\dot{v}_{y}=&\left[ y-2v_{x}-y\left(\frac{\mu_{1}}{d_{1}^{3}}+\frac{\mu_{2}}{d_{2}^{3}}\right) \right] \\
&+2\omega_{1}\left( {y} - v_{x} \right)/c^{2} + \mathcal{Q}(x,y,v_x,v_y)/(ac^{2}).\\
\end{aligned}\end{equation}\\
where $\mathcal{P}(x,y,v_x,v_y) $ and  $\mathcal{Q}(x,y,v_x,v_y) $
are obtained from $\mathcal{L}_{2}$. Note that $1/(c^2a)$ is the
1PN effect. {Here, the speed of light $c$ is not
necessarily set to the geometric unit 1 although $G$ takes the
geometric unit 1. Dubeibe et al. [31] founded that the case of
$a=1$ and $c=10^4$ corresponds to the main relativistic effects
with an order of $10^{-8}$ in the solar system. If $a=c=1$, the
main relativistic effects $\sim 1$ are relatively poor PN
approximations. In this case, the approximate Euler-Lagrange
equations fail to conserve the Jacobi integral, as was reported by
Dubeibe et al. Of course, $c=1$ is often used in strong
gravitational fields of compact objects. To make the PN
approximations valid, one should give $c$ a larger value for
$a=1$, or $a$ a larger value for $c=1$ [59, 60]. In fact, $c$ has
different values in different unit systems [61]. Thus, we take $c$
as a free parameter in the following numerical simulations in
Sections III. B and C so that the PN approximations $1/(c^2a)$
remain useful. }

\subsection{Numerical evaluations}

An eighth- and ninth-order Runge-Kutta-Fehlberg integrator [RKF89]
with adaptive step sizes is used to provide high-precision
reference solutions for evaluating the numerical performance of
low order methods such as EM4. With the aid of this integrator,
the solutions of the exact Euler-Lagrange equations (11), (12),
(29) and (30) and the approximate Euler-Lagrange equations (32)
can be obtained. In this way, phase-space structures of four
orbits in the two sets of equations are described on the
Poincar\'{e} section $y=0$ with $v_y>0$ in Fig. 1. Here, the
initial conditions are $y=v_{x}=0$, and the parameters are
$\mu_{2}=0.001$, $\mu_{1}=1-\mu_{2}$ and $C_{j}=3.07$. As is
above-mentioned, $c$ is used as a free parameter. When the
parameters $a$ and $c$ and the starting value of $x$ are
considered, the initial value of $v_y>0$ is solved from Eq. (31).

The two sets of equations have almost the same
Kolmogorov-Arnold-Moser (KAM) torus for the starting value
$x=0.35$ with parameters $c=100$ and $a=1$ in Fig. 1(a). This KAM
torus indicates the regularity of the integrated orbit. Although
the two sets of equations yield regular KAM tori for the starting
value $x=0.28$ with parameters $a=10$ and $c=100$ in Fig. 1(b),
the two tori are typically different. For the initial value
$x=0.4895$ with parameters $c=100$ and $a=1$ in Fig. 1(c), the
exact  equations and the approximate ones produce approximately
same chaotic solutions with  many points filled with areas. For
the initial value $x=0.687$ with $c=100$ and $a=2$ in  Fig. 1(d),
chaos occurs in the approximate equations, whereas does not in the
exact equations. Figs. 1 (b) and (d) show that the approximate
equations and the exact ones have distinct phase-space structures,
i.e., distinct solutions. This supports the results of [29, 30]
again. Which of the approximate Euler-Lagrange equations and the
exact ones can give correct solutions to the Lagrangian (25)? The
exact Euler-Lagrange equations can without question. Thus, they
are used as a test model in the following numerical simulations.

RKF89 can significantly improve the accuracy, as can be verified
by comparison with lower-order integrators like EM4.  However, the
improved accuracy requires that RKF89 be more computationally
demanding than these lower-order integrators for long-term
integrations. If the lower-order integrators can provide reliable
results, they should be employed to study the trajectories and to
detect the chaotical behavior. In view of this point, three
fourth-order integrators are  compared with EM4. They are an
implicit symplectic method (IM4) consisting of three second-order
implicit midpoint rules [35], a Runge-Kutta integrator (RK4), and
a Gauss-Runge-Kutta (GRK) implicit symplectic method [34, 37-39].
The four methods EM4, IM4, GRK and RK4 can yield the same
phase-space structures to the four orbits in Fig. 1 for short
integration times. When the integrations are long enough, the
three methods EM4, IM4 and GRK still have the same phase-space
structures as those of Fig. 1, but RK4 does not have.

When the step size is chosen as $h=0.01$, Figs. 2 (a) and (b) plot
relative energy errors of the regular orbit in Fig. 1(a) and the
chaotic orbit in Fig. 1(c). The relative energy errors are
calculated by $\Delta E=|(E_t-E_0)/E_0|$, where $E_0$ is the
energy in Eq. (4) at time 0 and $E_t$ stands for the energy
calculated at integration time $t$. The errors remain bounded for
the three geometric integrators EM4, IM4 and GRK, whereas grow
with time for RK4. They are about orders of $10^{-12}$ to
$10^{-6}$ for EM4, IM4 and GRK, and $10^{-12}$ to $10^{-5}$ for
RK4. The accuracies are good and acceptable in the machine's
double precision. The methods with the errors from small to large
are GRK, EM4, IM4 and RK4. The calculations in the chaotic case
(Fig. 2(b))  appear to be more accurate than in the regular ones
(Fig. 2(a)). {An explanation to this result is
given here. Based on the theory of numerical calculations, the
truncation error in energy is approximately estimated by $(h/T)^k$
for a $k$th-order symplectic method, and it in solutions is
estimated by $(h/T)^{k+1}$, where $T$ is an orbital period. The
period for the ordered orbit in Fig. 2(a) is $T\approx 5.11$.
Although the orbit is chaotic in Fig. 2(b), it has an approximate
average period $T\approx 7.91$. For a given time step, the energy
accuracies get higher with an increase of the orbital periods.
Thus, it is reasonable that the energy accuracies for the regular
case in Fig. 2(a) are poorer than those for the chaotic case in
Fig. 2(b). However, the solutions' accuracies should be better for
former case than those for the latter case because the solutions
in the chaotic case exhibit exponentially sensitive dependence on
the initial conditions. When a smaller step size $h=0.001$ is used
in Figs. 2 (c) and (d), the energies have higher accuracies and
are accurate to orders of $10^{-15}\sim 10^{-11}$ for the three
geometric integrators. To ensure such higher enough accuracies, we
adopt the smaller step size $h=0.001$ in the PN circular
restricted three-body problem. }

It is worth noticing the differences in computations among the
three geometric methods EM4, IM4 and GRK. The position
$\textbf{r}$ and velocity $\textbf{v}$ must be solved in terms of
the iteration method in IM4 and GRK, but only the solutions of
$\textbf{v}$ need iterations in EM4. This seems to show that the
computational cost for EM4 is less than for IM4 or GRK. To check
the result, we plot computational efficiencies in Fig. 3. The test
orbit is that of Fig. 1(a). A series of step sizes $h=10^{-3}$,
$10^{-3}\times1.2$, $10^{-3}\times1.2^{2}$,
$10^{-3}\times1.2^{3}$, $\cdots$  are considered. When one of the
time steps is given, the maximum relative energy errors for the
three schemes during the integration time $t=10^5$ are drawn in
Fig. 3(a). GRK exhibits the best accuracy. EM4 and IM4 are almost
the same in the accuracies. Each of the maximum errors
corresponding to CPU time is shown in Fig. 3(b). When the three
methods use different time steps and provide the same accuracy,
EM4 or GRK takes less CPU time than IM4. EM4 and GRK take almost
the same CPU times. Given CPU time, the accuracy of EM4 is two or
three orders of magnitude better than that of IM4, and is
approximately the same as that of GRK. Obviously, the efficiency
of EM4 is superior to that of IM4. In other words, EM4 needs less
computational cost than IM4 and GRK when the time step and
integration time are fixed.

\subsection{Dependence of chaos on the initial conditions and parameters}

Considering reliable computational accuracy and high efficiency
for given time step and integration time, we use EM4 to study the
orbital dynamical behavior of orbits in the exact equations. Apart
from the technique of Poincar\'{e} sections, Lyapunov exponents
[62] and fast Lyapunov indicators (FLIs) [63-65] are often used to
distinguish between the regularity and chaoticity of bounded
orbits. The FLI is defined as
\begin{equation}\label{Eq:fli}
FLI=\log_{10}\frac{d(t)}{d(0)},
\end{equation}
where $d(t)$ and $d(0)$ are the distances between two nearby
trajectories at times $t$ and 0. The initial separation $d(0)=
10^{-9}$ is a {sufficient} choice. The FLI with
two nearby trajectories proposed in Ref. [63] is originated from a
modified version of the FLI with two tangent vectors given in
Refs. [64, 65], and is also from a modified version of the
Lyapunov exponents with two nearby trajectories [62]. For
convenience of analytical discussion, the original FLI (denoted by
$\delta=\ln(d(t)/d(0))$) is adopted. If $\delta=e^{\ln t}$, then
$d(t)=d(0)e^{t}$. This indicates that the distance $d(t)$ grows
exponentially with time $t$. This characteristic is just the
description of the largest Lyapunov exponent, which shows the
chaoticity of a bounded orbit. If $\delta=k\ln t$, then
$d(t)=d(0)t^k$. This indicates that the distance $d(t)$ grows in a
power law of time $t$. It is the characteristic of a regular
bounded orbit. { There are two particular cases
about the relation $\delta=k\ln t$. When $k=1$, $\delta= \ln t$
corresponds to regular orbits. Lukes-Gerakopoulos et al. [67] have
shown that $\delta$ behaves like $\delta= k \ln t$ for a transient
time before the exponential growth of weak chaotic orbits takes
over. }  An exponential increase of the distance is much larger
than a polynomial increase of the distance. Such typically
different ratios of growth of FLIs with time are used to detect
chaos from order. An exponential growth of FLI with time
$\log_{10} t$ indicates the chaoticity of a bounded orbit, but a
linear growth of FLI with time shows the regularity of a bounded
orbit.

The FLIs of two orbits with initial values $x=0.4$ and $x=0.28$
are shown in Fig. 4. The parameters and other initial conditions
(except the initial value $v_y$) are those of Fig. 1(a). The
initial value $x=0.4$ corresponds to the regularity, and the
initial value $x=0.28$ indicates the chaoticity. It is found that
7.5 is a threshold value of FLIs between the regular and chaotic
cases when the integration time reaches $t=10^4$. The FLIs larger
than 7.5 indicate the presence of chaos, whereas the FLIs no more
than the threshold mean the presence of order. The technique of
FLIs is convenient to trace a transition from order to chaos with
a parameter or an initial condition varying. Fig. 5 draws the
dependence of FLIs on the initial values $x$. The parameters and
other initial conditions (except the initial value $v_y$) are the
same as those of Fig. 1(a). The initial values $x$ corresponding
to order and chaos are shown clearly. Two larger regular intervals
for the onset of order are $0.346\leq x\leq 0.485$  and $0.667\leq
x\leq 0.805$. There are four larger chaotic intervals: $0.254\leq
x\leq 0.295$, $0.3125\leq x\leq 0.3455$, $0.463\leq x\leq
0.56185$, and $0.5545\leq x\leq 0.6335$. A number of smaller
regular or chaotic intervals are also present. Taking the FLIs
neighboring $x=0.666$ as examples, we focus on the transition
between order and chaos. FLI=5.3 for $x=0.665$  and FLI=5.2 for
$x=0.667$ correspond to the regular case, but FLI=21.1 for
$x=0.666$ corresponds to the chaotic case. The regularity and
chaoticity are shown through the Poincar\'{e} sections in Fig.
5(b).

Fig. 6(a) describes the dependence of FLIs on parameter $a$ and
the initial value $x$. Given  $c=10^4$, $a$ runs from 1 to 10 with
an interval of 0.1, and $x$ ranges  from 0.1 to 0.9 with an
interval of 0.01. The other parameters and initial conditions
(except the initial value $v_y$) are still those of Fig. 1(a). A
great two-dimensional space of  $a$ and $x$ corresponds to the
regularity of bounded orbits. There are larger unstable regions
colored Black. Many smaller areas for the chaoticity of bounded
orbits exist. Chaos mainly occurs in the neighbourhood of $x=0.1$
or $a=1$. Of course, chaos also easily appears in the boundary
regions between the ordered and unstable regions. For example, the
orbit is ordered for $a=1$ and $x=0.815$ with FLI=5.59, while
chaotic for $a=1.1$ and $x=0.815$ with FLI=46.88, as can be seen
from the Poincar\'{e} sections in Fig. 6(b). In Fig. 6(c) with
$a=1$, $c$ ranges from 100 to 10000, and $x$ runs from 0.1 to 0.9.
There are many larger two-dimensional spaces of $c$ and $x$ for
the onset of order and chaos. Thinner unstable regions can be met
in the neighbourhood of $x=0.1$ or $x=0.9$. Fig. 6(d) shows that
order exists for $x=0.632$ and $c=190.5$ with FLI=5.91, while
chaotic for $x=0.632$ and $c=208.9$ with FLI=12.48.

When $a=1$ and $c=10^4$ are given in Fig. 7(a), the Jacobi
constant $C_j$ runs from 2 to 8 with an interval of 0.06, and the
initial position $x$ is also considered from 0.1 to 0.9. The other
parameters and initial positions are those of Fig. 1(a). The
values of $C_j$ and $x$ are more for the regular case than for the
chaotic case, whereas less than for the unstable case. Fig. 7(b)
displays that the orbit is ordered for $C_j=3.125$ and $x=0.52875$
with FLI=4.89, while chaotic for $C_j=3.0875$ and $x=0.52875$ with
FLI=111.06.

Two points are illustrated here. \textcolor{blue}{The} dynamical
results obtained from EM4 in Figs. 4-7 are consistent with those
given by RKF89. In addition, the technique of FLIs can sensitively
distinguish between the regular and chaotic two cases. It is
effective to trace the transition from order to chaos with a
variation of one or two initial conditions and parameters in Figs.
5-7.

\section{Spinning compact binaries}

A PN Lagrangian system of two spinning black holes is introduced
simply in Sect. 4.1. When the spins are expressed in a set of
canonical coordinates, the symplectic-like integrator EM4 is
applied to the exact Euler-Lagrange equations and its performance
is evaluated in Sect. 4.2. The orbital dynamics of order and chaos
in the exact Euler-Lagrange equations are investigated in Sect.
4.3.

\subsection{Dynamical equations}

Let us consider two black holes with masses $M_{1}$ and $M_{2}$.
They have the total mass $M=M_{1}+M_{2}$, the reduced mass
$\mu=M_{1}M_{2}/M $, and $\nu=\mu/M $. Body 2 relative to body 1
has position $\mathbf{r}$ and velocity $\mathbf{v}$. We take
$r=|\mathbf{r}|$, $\mathbf{n}=\mathbf{r}/r$, and
$\dot{r}=\mathbf{n}\cdot\mathbf{v}$. The two bodies evolve
according to the PN Lagrangian in the ADM coordinates [56, 66]:
\begin{equation}\label{Eq:cb-L}
\mathcal{L}(\mathbf{r}, \mathbf{v}, \mathbf{S}_1,
\mathbf{S}_2)=\mathcal{L}_{O}+\mathcal{L}_S.
\end{equation}
$\mathcal{L}_{O}$ is an orbital part with the following expression
\begin{equation}\label{Eq:cb-Lo}
\mathcal{L}_{O}=\mathcal{L}_{N}+\frac{1}{c^{2}}\mathcal{L}_{1PN}+\frac{1}{c^{4}}
\mathcal{L}_{2PN},
\end{equation}
where the Newtonian term $\mathcal{L}_{N}$ and the 1PN and 2PN
contributions $\mathcal{L}_{1PN}$ and $\mathcal{L}_{2PN}$ are
\begin{equation}\label{Eq:cb-LN}
\mathcal{L}_{N}=\frac{1}{r}+\frac{\mathbf{v}^{2}}{2},
\end{equation}
\begin{equation}\label{Eq:cb-L1NP}\begin{aligned}
\mathcal{L}_{1PN}=&\frac{\mathbf{v}^{4}}{8}-\frac{3 \nu
\mathbf{v}^{4}}{8}-\frac{1}{2 r^{2}}
\\
& +\frac{1}{r}\left(\frac{\nu {\dot{r}}^{2}}{2}+\frac{3
\mathbf{v}^{2}}{2} +\frac{\nu \mathbf{v}^{2}}{2}\right),
\end{aligned}\end{equation}
\begin{equation}\label{Eq:cb-L2NP}\begin{aligned}
\mathcal{L}_{2PN}=&\frac{\mathbf{v}^6}{16}-\frac{7\nu\mathbf{v}^{6}}{16}+\frac{13\nu^{2}\mathbf{v}^6}{16}\\
&+\frac{1}{r}\left( \frac{3\nu^{2}\dot{r}^4}{8} + \frac{\nu\dot{r}^{2}\mathbf{v}^{2}}{2}
-\frac{5\nu^{2}\dot{r}^{2}\mathbf{v}^{2}}{4}\right.\\ &\left.
+\frac{7\mathbf{v}^{4}}{8}-\frac{3\nu\mathbf{v}^{4}}{2}-\frac{9\nu^2\mathbf{v}^{4}}{8}\right)\\
&+\frac{1}{r^{2}}\left(\frac{3\nu\dot{r}^{2}}{2} + \frac{3\nu^2\dot{r}^{2}}{2} +2\mathbf{v}^{2}
-\nu\mathbf{v}^{2}+\frac{\nu^{2}\mathbf{v}^{2}}{2}\right)\\
&+\frac{1}{r^{3}}\left(\frac{1}{4}+\frac{3\nu}{4}\right).
\end{aligned}\end{equation}
The two bodies spin according to the spin effects
\begin{equation}
\mathcal{L}_{S}=\frac{1}{c^3}\mathcal{L}_{1.5SO}+\frac{1}{c^4}\mathcal{L}_{2SS},
\end{equation}
where the 1.5PN spin-orbit coupling $\mathcal{L}_{1.5SO}$ and the
2PN spin-spin effect $\mathcal{L}_{2SS}$ [26] are
\begin{equation}\label{Eq:cb-Lso}
\mathcal{L}_{1.5SO}=\frac{\nu}{r^{3}}\mathbf{v}\cdot\left[\mathbf{r}\times\left(\gamma_{1}\mathbf{S_1}+\gamma_{2}\mathbf{S_2}\right)\right],
\end{equation}
\begin{equation}\label{Eq:cb-Lss}
\mathcal{L}_{2SS}=-\frac{\nu}{2r^{3}}\left[\frac{3}{r^{3}}\left(\mathbf{S_{0}}\cdot\mathbf{r}\right)^{2}
- \mathbf{S_{0}}^{2}  \right].
\end{equation}
Note that $\gamma_{1}=2+3/(2\beta)$, $\gamma_{2}=2+3\beta/2$, and
$\mathbf{S_{0}}=\left(1+1/\beta \right)\mathbf{S_1}+\left(1+\beta
\right)\mathbf{S_2} $, where the mass ratio is
$\beta=M_{2}/M_{1}$. Eqs. (34)-(41) are dimensionless through a
series of scale transformations:  $\mathbf{r}\rightarrow
GM\mathbf{r}$, $t\rightarrow GMt$, $\mathbf{S_i}\rightarrow G \mu
M\mathbf{S_i} ~(i=1,2)$, and $\mathcal{L}\rightarrow
\mu\mathcal{L}$. $G$ uses a geometric unit $G=1$.

The Hamiltonian (7) for the PN Lagrangian (34) reads
\begin{equation}
\mathcal{H}(\mathbf{r}, \mathbf{v}, \mathbf{S}_1,
\mathbf{S}_2)=\mathbf{v}\cdot \mathbf{p}-\mathcal{L}(\mathbf{r},
\mathbf{v}, \mathbf{S}_1, \mathbf{S}_2),
\end{equation}
where the momentum $\mathbf{p}$ is still defined in Eq. (1) by
\begin{equation}
\mathbf{p}=\frac{\partial}{\partial
\mathbf{v}}\mathcal{L}(\mathbf{r}, \mathbf{v}, \mathbf{S}_1,
\mathbf{S}_2).
\end{equation}
{As is aforementioned, $\mathbf{v}$ in Eq. (43)
can be expressed as a function of $\mathbf{r}$, $\mathbf{p}$,
$\mathbf{S}_1$ and $\mathbf{S}_2$. The higher-order terms are
truncated in general if $\mathbf{v}$ is expressed in terms of
$\mathbf{p}$. However, $\mathbf{v}$ in Eq. (42) has no such
operation and is still remained. This means that Eq. (42) is only
a formal Hamiltonian but is not a standard Hamiltonian (note that
the standard Hamiltonian should be a function of coordinates and
momenta rather than a function of coordinates and velocities).
That is to say, the formal Hamiltonian, as a function of
$\mathbf{r}$, $\mathbf{v}$, $\mathbf{p}$, $\mathbf{S}_1$ and
$\mathbf{S}_2$, comes directly from the Legendre transformation of
the Lagrangian and has no terms truncated. Thus, it is exactly
equivalent to this Lagrangian.} The PN Lagrangian (34) has four
integrals of motion, which involve the energy integral
\begin{equation}
E=\mathcal{H}(\mathbf{r}, \mathbf{v}, \mathbf{S}_1, \mathbf{S}_2),
\end{equation}
and the total angular momenta
\begin{equation}
\mathbf{J}=\mathbf{r}\times \mathbf{p}+\mathbf{S}_1+ \mathbf{S}_2.
\end{equation}
{Because no terms are truncated when the formal
Hamiltonian is obtained from the Lagrangian, the energy is an
exact integral in the Lagrangian.}

Apart from the four integrals, the two spin lengths
$S_{i}=\chi_{i} M^2_{i}/(\mu M)~(0\leq\chi_{i}\leq 1)$ also remain
constant. Using the constant spin lengths, Wu $\&$ Xie [33]
introduced the canonical spin coordinates $(\theta_{i},\xi_{i})$
as follows:
\begin{equation}
 \boldsymbol{S}_{i}=\left(\begin{array}{l}
\rho_{i} \cos \theta_{i} \\
\rho_{i} \sin \theta_{i} \\
\xi_{i}
\end{array}\right),
\end{equation}
where $\rho_{i}$ are expressed as
\begin{equation}\label{Eq:spin-var3}
\rho_{i}=\sqrt{S^2_{i}-\xi^2_{i}}.
\end{equation}
{Now, the Hamiltonian can be expressed in terms of
ten independent phase space variables, consisting of five degrees
of freedom (i.e., generalized coordinates) $(\mathbf{r}, \theta_1,
\theta_2)$ and five conjugate momenta $(\mathbf{p}, \xi_1,
\xi_2)$. These independent variables evolve by satisfying the
Hamiltonian canonical equations of motion:}
\begin{eqnarray}\label{Eq:spin-em}
\dot{\mathbf{r}}_i &=&\frac{\partial\mathcal{H} }{\partial \mathbf{p}}=\mathbf{v},\\
\dot{\theta}_i &=&\frac{\partial\mathcal{H} }{\partial \xi_i}
=-\frac{\partial \mathcal{L} }{\partial \mathbf{S}_i}\cdot \frac{\partial \mathbf{S}_i}{\partial \xi_i};\\
\dot{\mathbf{p}} &=&-\frac{\partial\mathcal{H} }{\partial
\mathbf{r}}
=\frac{\partial \mathcal{L} }{\partial \mathbf{r}},\\
\dot{\xi}_i &=&-\frac{\partial\mathcal{H} }{\partial
\mathbf{\theta}_i} =\frac{\partial \mathcal{L} }{\partial
\mathbf{S}_i}\cdot \frac{\partial \mathbf{S}_i}{\partial
\theta_i}.
\end{eqnarray}
{ $\mathbf{r}$ and $\mathbf{p}$ are a pair of
canonical variables. So are $\theta_i$ and $\xi_i$.} After the
solutions $(\textbf{r},\theta_{1},\theta_{2},
\textbf{p},\xi_{1},\xi_{2})$ are solved from Eqs. (48)-(51),
$\textbf{v}$ is calculated iteratively in terms of Eq. (43). Eqs.
(48)-(51) with Eq. (43) are strictly derived from the Lagrangian
(34) and exactly conserve the energy (44). In this case, the four
algorithms EM4, IM4, GRK and RK4 are available for the exact
Euler-Lagrange equations (48)-(51).
The solutions are not given analytically because the existence of the
four integrals including the total energy and the total momenta in
the 10-dimensional phase space determines the non-integrability of
the formal PN Hamiltonian (or Lagrangian) formulation.

Another consideration is that the total accelerations to the 2PN order can be derived from the Euler-Lagrange  equations of the Lagrangian (34)
and are written as
\begin{equation}\label{Eq:51} \begin{aligned}
\frac{d\mathbf{v}}{dt}=\frac{\partial
\mathcal{L}}{\partial\mathbf{r}}+\frac{\tilde{\mathbf{a}}_{1PN}}{c^{2}}
+\frac{\tilde{\mathbf{a}}_{1.5SO}}{c^{3}}+\frac{\tilde{\mathbf{a}}_{2PN}}{c^{4}}
+\cdots.
\end{aligned}\end{equation}
{  The PN accelerations such as
$\tilde{\mathbf{a}}_{1PN}$ in the right-hand side of Eq. (52) are
originated from the derivative of momenta (43) with respect to
time (i.e., $d\mathbf{p}/dt$) in the Euler-Lagrange equations.
Because the PN terms higher than the 2PN terms are dropped in the
total accelerations, Eq. (52) with Eqs. (48), (49) and (51) is the
approximate equations of the Lagrangian (34). Of course, such
approximate equations do not exactly conserve the energy (44).}

{It is worth pointing out that the differences
in the PN contributions (including the spin effects) between the
two sets of equations are apparent. To show the differences, we
rewrite Eqs. (43) and (5) of the exact equations as
\begin{eqnarray}
\mathbf{\dot{v}} &=& \frac{\partial \mathcal{L}}{\partial
\mathbf{r}}+\mathbf{f}_{1PN}(\mathbf{r},\mathbf{v},\mathbf{\dot{v}})
+\mathbf{f}_{2PN}(\mathbf{r},\mathbf{v},\mathbf{\dot{v}})
\nonumber \\ &&
+\mathbf{f}_{SO}(\mathbf{r},\mathbf{v},\mathbf{S}_1, \mathbf{S}_2,
\mathbf{\dot{S}}_1, \mathbf{\dot{S}}_2),
\end{eqnarray}
where $\mathbf{f}_{1PN}$, $\mathbf{f}_{2PN}$ and $\mathbf{f}_{SO}$
are functions associated to the PN accelerations. This equation is
an implicit equation with respect to the acceleration
$\mathbf{\dot{v}}$. When $\mathbf{\dot{v}}=-\mathbf{r}/r^3$ (i.e.,
the Newtonian acceleration is given to $\mathbf{\dot{v}}$) and
$\mathbf{\dot{S}}_1=\mathbf{\dot{S}}_2=0$ in the right-hand side
of Eq. (53), the approximate equations (52) is obtained. However,
these approximations are not given to the exact equations. If
$\mathbf{\dot{v}}$ takes the 1PN orbital contribution, 1.5PN
spin-orbit term, 2PN orbital term and 2PN spin-spin coupling in
the function $\mathbf{f}_{1PN}$, then $\mathbf{f}_{1PN}$ includes
the 2PN orbital contribution, 2.5PN spin-orbit term, 3PN orbital
term and 3.5PN spin-orbit coupling. In this case,
$\mathbf{f}_{2PN}$ contains the 3PN orbital contribution, 3PN
spin-spin effect, 4PN orbital term and 4PN spin-spin coupling. If
$\mathbf{\dot{S}}_1$ and $\mathbf{\dot{S}}_2$ in  the function
$\mathbf{f}_{SO}$ are Eqs. (49) and (51), then $\mathbf{f}_{SO}$
has 3PN spin-spin coupling and 3.5PN spin-orbit interaction. Thus,
besides the terms in the approximate equations (52), many other
terms such as the 2.5PN spin-orbit coupling and the 3PN spin-spin
contribution are included in the exact equations (53). The 2.5PN
spin-orbit and 3PN spin-spin contributions are implicitly hidden
in the exact equations, but are absent in the approximate
equations.

\subsection{Numerical tests}

Same as $G$, the speed of light $c$ also takes the geometric unit
$c=1$. The parameters are given by $\chi_{1}=\chi_{2}=1$ and
$\beta=4$. The initial conditions are chosen as $x=70$, $y=0$,
$z=0$, $p_{x}=p_{z}=0$, and $p_{y}=\sqrt{(1-e)/x} $ with initial
eccentricity $e=0.16$. The initial canonical spin variables are
given by $\theta_{1}=\theta_{2}=\pi/2$, $\xi_{1}=0.1$ and
$\xi_{2}=0.95$. When RKF89 is used, the approximate Euler-Lagrange
equations and the exact ones have different three-dimensional
orbits after a time of integration in Fig. 8. As is mentioned
above, the exact Euler-Lagrange equations should be chosen as the
equations of motion for the Lagrangian (34).

Now, RKF89 is replaced with one of the four algorithms EM4, IM4,
GRK and RK4. The step size $h=1$ is fixed. Fig. 9 plots the
relative energy errors $\Delta E$, relative total angular momentum
errors $\Delta J$ and relative position errors $\Delta R$ for the
four integrators. Here, $\Delta E=|(E_t-E_0)/E_0|$, and $\Delta
J=|\textbf{J}_t-\textbf{J}_0|/|\textbf{J}_0|$, where
$\textbf{J}_0$ is the total angular momenta (45) at time 0 and
$\textbf{J}_t$ denotes the total angular momenta calculated at
integration time $t$. The positions calculated by these methods at
time $t$ are $\textbf{r}_E$ for EM4, $\textbf{r}_I$ for IM4,
$\textbf{r}_G$ for GRK, and $\textbf{r}_R$ for RK4. The
higher-precision method RKF89 is used to provide reference
solutions such as position $\textbf{r}_F$. In this way, the
relative position errors between RKF89 and EM4 can be computed by
$\Delta R=|\textbf{r}_F-\textbf{r}_E|/|\textbf{\textbf{r}}_F|$.
The position errors between RKF89 and one of IM4, GRK and RK4 are
also calculated in this similar way.  The three geometric methods
EM4, IM4 and GRK give no secular growth to the errors in the total
energy and the  total angular momenta. In other words, they
conserve the total energy and the total angular momenta. They have
no dramatic but minor differences in the accuracies, and their
accuracies are almost approximate to the machine's precision. In
the energy accuracies, GRK is slightly better than EM4 or IM4; EM4
and IM4 are approximately the same. In the angular momentum
accuracies, EM4 and GRK are basically the same, and are slightly
superior to IM4. However, RK4 does not conserve the total energy
and the total angular momenta, and exhibits the poorest
accuracies. The three methods EM4, IM4 and GRK are almost the same
in the relative position errors as the integration lasts. The
position errors for EM4, IM4 and GRK are several orders of
magnitude smaller than those for RK4.}

The numerical tests show that the three geometric integrators
exhibit good long-term performance in the conservation of energy
and angular momentum, but RK4 exhibits poor long-term performance.
For given time step and integration time, EM4 is superior to IM4
and GRK in computational efficiency.

\subsection{Orbital dynamics}

EM4 combined with  the technique of FLIs is employed to explore
the dynamical behavior of orbits. The initial conditions are the
same as those in Fig. 9, except for the initial eccentricity
$e=0.66$ and the initial spin angles $\theta_{1}$ and
$\theta_{2}$. The initial spin angles $\theta_{1}=5.4349$ and
$\theta_{2}=2.042$ with FLI= 2.56 yield a regular solution in Fig.
10. However, the initial spin angles $\theta_{1}=4.7123$ and
$\theta_{2}=2.4504$ with FLI= 10.48 exhibit a chaotic solution.
The FLIs larger than 7.5 indicate the chaoticity, whereas the FLIs
no more than than 7.5 indicate the regularity when the integration
time reaches $t=10^5$. The performance of this integrator in the
conservation of energy and angular momenta is independent of the
regularity or chaoticity of orbits.

Using the FLIs, we can classify the two-dimensional space of
initial spin angles according to different orbital dynamical
features in Fig. 11(a). Our method is described here. The initial
conditions and parameters are the same as those in Fig. 10. The
initial spin angles $\theta_1$ and $\theta_2$ range from 0 to
$2\pi$ with an interval of $0.01\times \pi$. The FLI for each pair
of $\theta_1$ and $\theta_2$ is obtained after the integration
time $t=10^5$. Based on the FLIs, the initial spin angles
$\theta_1$ and $\theta_2$ can be divided into three regions:
regular region of bounded orbits, chaotic region of bounded orbits
and unstable region. A number of initial spin angles correspond to
the regularity, and many initial spin angles colored Red indicate
the presence of chaos. There are a lot of initial spin angles
colored Black leading to the instability. No rule on  the
transition from order to chaos  can be given as the initial spin
angles are varied. Fig. 11(b) relates to the dependence of FLIs on
the initial eccentricities $e$ and initial separations  $x$ with
mass ratio $\beta = 3$. The other initial conditions and
parameters are the same as those of Fig. 9. There are two main
parts. The largest region colored Blue corresponds to order. The
larger region colored Black shows that none of the orbits is
stable below the horizon line $x=21.8$. Only a small number of
values of $x$ and $e$ on the boundary between order and
instability indicate the onset of chaos. For instance, the orbit
with $e=0.75$ and $x=65$ having FLI=12.3 is chaotic. The result
concluded in Fig. 11(b) is that the orbits become unstable as the
initial eccentricities increase or the initial separations
decrease, whereas stable and regular as the initial separations
increase.

Some explanations to the above results are given here. The exact
Euler-Lagrange equations are conservative, integrable and
nonchaotic when the two bodies do not spin. However, when the two
bodies are spinning, the spin contributions lead to the
nonintegrability and probable chaoticity of  the exact equations.
The spin effects play an important role in the onset of chaos. As
is claimed above, the 1.5PN spin-orbit and 2PN spin-spin
contributions are explicitly included in the exact equations, and
the 2.5PN spin-orbit and 3PN spin-spin contributions are
implicitly hidden. The 2.5PN spin-orbit and 3PN spin-spin
contributions hidden in the exact equations  would easily induce
the instability or chaoticity of the solutions for the exact
equations. Larger initial eccentricities and smaller initial
separations  $x$ cause an increase of the spin effects, and
therefore chaos or instability is easily induced.

\section{Conclusion}

A PN Lagrangian formalism can be exactly equivalent to the same
order PN \emph{formal} Hamiltonian \textcolor{blue}{(7)}, where
the positions and momenta are phase-space variables but velocities
are implicit functions of the positions and momenta. Such a formal
Hamiltonian seems to explicitly depend on the positions,
velocities and momenta. Because these velocities are solved
iteratively from the algebraic equations of the momenta defined by
the Lagrangian, this Hamiltonian is still a function of the
positions and momenta. {The Lagrangian formalism
is not exactly equivalent to the same order PN standard
Hamiltonian (5), which is a function of the positions and momenta.
} The canonical equations for {the formal PN
Hamiltonian (7)} must strictly conserve the Hamiltonian quantity,
i.e., energy. They are the coherent or exact PN Euler-Lagrange
equations of motion.

Doubling the phase-space variables including the  positions and
momenta in the formal Hamiltonian, we introduce a new Hamiltonian
in extended phase space. This new Hamiltonian consists of two
parts: one of which is equal to the original formal Hamiltonian
depending on the original positions and the new momenta with the
new velocities, and another of which is  equal to the original
Hamiltonian depending on the original momenta and the new
positions with original velocities. The solutions of Hamilton's
canonical equations for the two parts of the new Hamiltonian are
used to design the standard second-order symplectic leapfrog
methods and fourth-order symplectic schemes. When these algorithms
are combined with the midpoint permutations, the extended
phase-space symplectic-like integrators become easily available
for the coherent PN Euler-Lagrange equations. In the course of
numerical integrations, the velocities must be solved iteratively.

Numerical tests show that a fourth-order extended phase-space
symplectic-like method exhibits good long-term stabilizing error
behavior in energy or angular momentum, as fourth-order implicit
symplectic method and Gauss-Runge-Kutta scheme do. The former
method takes less computational cost than the latter integrators
for given time step and integration time. This good numerical
performance of the extended phase-space method is independent of
the regularity or chaoticity of orbits. Because of such good
performance, the extended phase-space symplectic-like method with
the technique of FLIs is well applicable to studying the effects
of the parameters and initial conditions on the orbital dynamics
of the coherent Euler-Lagrange equations for a post-Newtonian
circular restricted three-body problem. The parameters and initial
conditions corresponding to order, chaos and instability are
found. They are also applied to trace the effects of the initial
spin angles, initial separations and initial orbital
eccentricities on the dynamics of the coherent post-Newtonian
Euler-Lagrange equations of spinning compact binaries.  As a
result, the initial spin angles, initial separations and initial
orbital eccentricities for the presence of order, chaos and
instability are obtained. Larger initial eccentricities and
smaller initial separations easily induce the occurrence of chaos
or instability.

\section*{Acknowledgments}

The authors are very grateful to a referee for useful suggestions.
This research has been supported by the National Natural Science
Foundation of China [Grant Nos. 11973020 (C0035736), 11533004,
11663005, 11533003, and 11851304], the Special Funding for Guangxi
Distinguished Professors (2017AD22006), and the National Natural
Science Foundation of Guangxi (Nos. 2018GXNSFGA281007 and
2019JJD110006).
%\newpage

%\bibliographystyle{apsrev4-1}
%\bibliography{library}

\begin{thebibliography}{99}
\addtolength{\itemsep}{+0.5 em}
%\setlength{\itemsep}{-5pt}

\bibitem{1} B. P. Abbott, et al., Phys. Rev. Lett. \textbf{116}, 061102 (2016).

\bibitem{2} R. Abbott, et al., Phys. Rev. Lett. \textbf{125}, 101102 (2020).

\bibitem{3} V. de Luca, V. Desjacques, et al. Phys. Rev. Lett. \textbf{126}, 051101 (2021).

\bibitem{4} J. Levin, Phys. Rev. Lett. \textbf{84}, 3515 (2000).

\bibitem{5} N. J. Cornish and J. Levin, Phys. Rev. Lett. \textbf{89}, 179001 (2002).

\bibitem{6} J. D. Schnittman and F. A. Rasio, Phys. Rev. Lett. \textbf{87}, 121101
(2001).

\bibitem{7} N. J. Cornish and J. Levin, Phys. Rev. D \textbf{68}, 024004
(2003).

\bibitem{8} M. D. Hartl and A. Buonanno, Phys. Rev. D \textbf{71}, 024027 (2005).

\bibitem{9} J. Levin, Phys. Rev. D. \textbf{74}, 124027 (2006).

\bibitem{10} X. Wu, Y. Xie, Phys. Rev. D \textbf{76}, 124004
(2007).

\bibitem{11} X. Wu and Y. Xie, Phys. Rev. D  \textbf{77}, 103012 (2008).

\bibitem{12} G. Huang, X. Ni, X. Wu, Eur. Phys. J. C \textbf{74}, 3012
(2014).

\bibitem{13} L. Huang, X. Wu and D. Zhu, Eur. Phys. J. C \textbf{76}, 488
(2016).

\bibitem{14} J. Levin, Phys. Rev. D \textbf{67}, 044013 (2003).

\bibitem{15} S. A. Hughes. Phys. Rev. Lett. \textbf{85}, 5480
(2000).

\bibitem{16} P. Jaranowski, and G. Sch\"{a}fer, Phys. Rev. D \textbf{57},
7274 (1998).

\bibitem{17} P. Jaranowski, and G. Sch\"{a}fer, Phys. Rev. D \textbf{60},
124003 (1999)

\bibitem{18} T. Damour, P. Jaranowski, and G. Sch\"{a}fer, Phys. Rev. D
\textbf{62}, 021501(R) (2000).


\bibitem{19} L. Blanchet, and G. Faye, Phys. Lett. A \textbf{271}, 58
(2000).

\bibitem{20} L. Blanchet, and G. Faye, Phys. Rev. D \textbf{63}, 062005
(2001).

\bibitem{21} L. Blanchet, and G. Faye, J. Math. Phys. \textbf{41}, 7675
(2000).

\bibitem{22} L. Blanchet, and G. Faye, J. Math. Phys. \textbf{42},
4391(2001).

\bibitem{23} V. C. de Andrade, L. Blanchet, and G. Faye, Class. Quantum
Grav. \textbf{18}, 753 (2001).


\bibitem{24} T. Damour, P. Jaranowski, and G. Sch\"{a}fer, Phys. Rev. D. \textbf{63}, 044021 (2001).

\bibitem{25} M. Levi and J. Steinhoff, J. Cosmol. Astropart. Phys. \textbf{12}, 003 (2014).

  %
\bibitem{26} X. Wu, L. Mei, G. Huang, and S. Liu, Phys. Rev. D \textbf{91}, 02042 (2015).

\bibitem{27} C. K\"{o}nigsd\"{o}rffer and A. Gopakumar, Phys. Rev. D \textbf{71}, 024039 (2005).

\bibitem{28}  A. Gopakumar and C. K\"{o}nigsd\"{o}rffer, Phys. Rev. D \textbf{72},
121501(R) (2005).

\bibitem{29} D. Li, X. Wu and E. Liang, Ann. Phys. \textbf{531}, 1900136 (2019).


\bibitem{30} D. Li, Y. Wang, C. Deng and X. Wu, Eur. Phys. J. Plus  \textbf{135}, 390 (2020).

\bibitem{31} F. L. Dubeibe, F. D. Lora-Clavijo, G. A. Gonzalez, Astrophys. Sp. Sci. \textbf{97}, 362 (2017).

%
\bibitem{32} X. Wu and G. Huang, Mon. Not. R. Astron. Soc. \textbf{452}, 3167 (2015).

\bibitem{33} X. Wu and Y. Xie, Phys. Rev. D \textbf{81}, 084045 (2010).

\bibitem{34} E. Hairer, C. Lubich and G. Wanner, 1999, Geometric Numerical
Integration (Berlin: Springer)

\bibitem{35} J. D. Brown,  Phys. Rev. D \textbf{73}, 024001 (2006).

\bibitem{36} O. Kop\'{a}\v{c}ek,V. Karas, J. Kov\'{a}\v{r}, Z. Stuchl\'{i}k, Astrophys. J., \textbf{722} , 1240 (2010).

\bibitem{37} J. Seyrich, G. Lukes-Gerakopoulos, Phys. Rev. D \textbf{86}, 124013
(2012).

\bibitem{38}  J. Seyrich, Phys. Rev. D \textbf{87}, 084064
(2013).

\bibitem{39} G. Lukes-Gerakopoulos,J. Seyrich,D. Kunst, Phys. Rev. D \textbf{90}, 104019
(2014).


\bibitem{40} S. Y. Zhong, X. Wu, S. Q. Liu, X. F. Deng, Phys. Rev. D \textbf{82}, 124040
(2010).

\bibitem{41} L. Mei, X. Wu and F. Liu,  Eur. Phys. J. C \textbf{73}, 2413 (2013).

\bibitem{42} L. Mei, M. Ju, X. Wu, S. Liu, Mon. Not. R. Astron. Soc. \textbf{435}, 2246
(2013).

\bibitem{43} L. Huang , L. Mei, Phys. Rev. D \textbf{100}, 24057 (2019).


\bibitem{44} Y. Wang, W. Sun, F. Liu and X. Wu, Astrophys. J. \textbf{907}, 66 (2021).

\bibitem{45} Y. Wang, W. Sun, F. Liu and X. Wu, Astrophys. J. \textbf{909}, 22 (2021).

\bibitem{46} Y. Wang, W. Sun, F. Liu and X. Wu, Astrophys. J. Supplement Series, \textbf{254}, 8 (2021).

\bibitem{47} X. Wu, Y. Wang, W. Sun and F. Liu, Astrophys. J. accepted (2021).

\bibitem{48} P. Pihajoki, Celest. Mech. Dyn. Astron. \textbf{121}, 211 (2015).

\bibitem{49} L. Liu, X. Wu, G. Q. Huang and F. Liu, Mon. Not. R. Astron. Soc. \textbf{459},
1968 (2016).

\bibitem{50} J. Luo, X. Wu, G. Huang, and F. Liu, Astrophys. J. \textbf{834}, 64 (2017).

\bibitem{51} D. Li and X. Wu, Mon. Not. R. Astron. Soc. \textbf{469}, 3031 (2017).

\bibitem{52} L. Liu, X. Wu and G. Q. Huang, Gen. Relativ. Gravit. \textbf{49}, 28
(2017).

\bibitem{53} J. Luo and X. Wu, Eur. Phys. J. Plus \textbf{132}, 485 (2017).

\bibitem{54} Y. Wu and X. Wu, International Journal of Modern Physics
C, \textbf{29}, 1850006 (2018).

\bibitem{55} G. Huang and X. Wu, Phys. Rev. D  \textbf{89}, 124034 (2014).

\bibitem{56} L. Blanchet, Living Rev. Relativ. \textbf{5}, 3 (2002).

\bibitem{57} H. Yoshida, Phys. Lett. A \textbf{150}, 262 (1990).

\bibitem{58} T. I. Maindl and R. Dvorak, Astron. Astrophys. \textbf{290}, 335 (1994).

\bibitem{59} X. Su, X. Wu,  F. Liu, Astrophys Space. Sci. \textbf{361}, 32 (2016).

\bibitem{60} R. Chen,X. Wu, Commun. Theor. Phys. \textbf{65}, 321 (2016)

\bibitem{61} J. Kla\v{c}ka , M. Kocifaj,  Mon. Not. R. Astron. Soc. \textbf{390}, 1491 (2008).

\bibitem{62} X. Wu and T. Y. Huang, Phys. Lett. A \textbf{313}, 77 (2003).

\bibitem{63} X. Wu, T. Y. Huang and H. Zhang, Phys. Rev. D \textbf{74}, 083001 (2006).

\bibitem{64} C. Froeschl\'{e}, E. Lega, and R. Gonczi, Celest. Mech. Dyn. Astron.
\textbf{67}, 41 (1997).

\bibitem{65} C. Froeschl\'{e} and E. Lega, Celest. Mech. Dyn. Astron. \textbf{78},
167 (2000).

\bibitem{66} L. Blanchet and B. R. Iyer, Class. Quantum Gravity \textbf{20}, 755.
(2003).

\bibitem{67} {G. Lukes-Gerakopoulos, N. Voligs, and C. Efthymiopoulos, Physica A \textbf{387}, 1907 (2008). }


\end{thebibliography}

\begin{figure*}[ptb]
\center{
\includegraphics[scale=0.25]{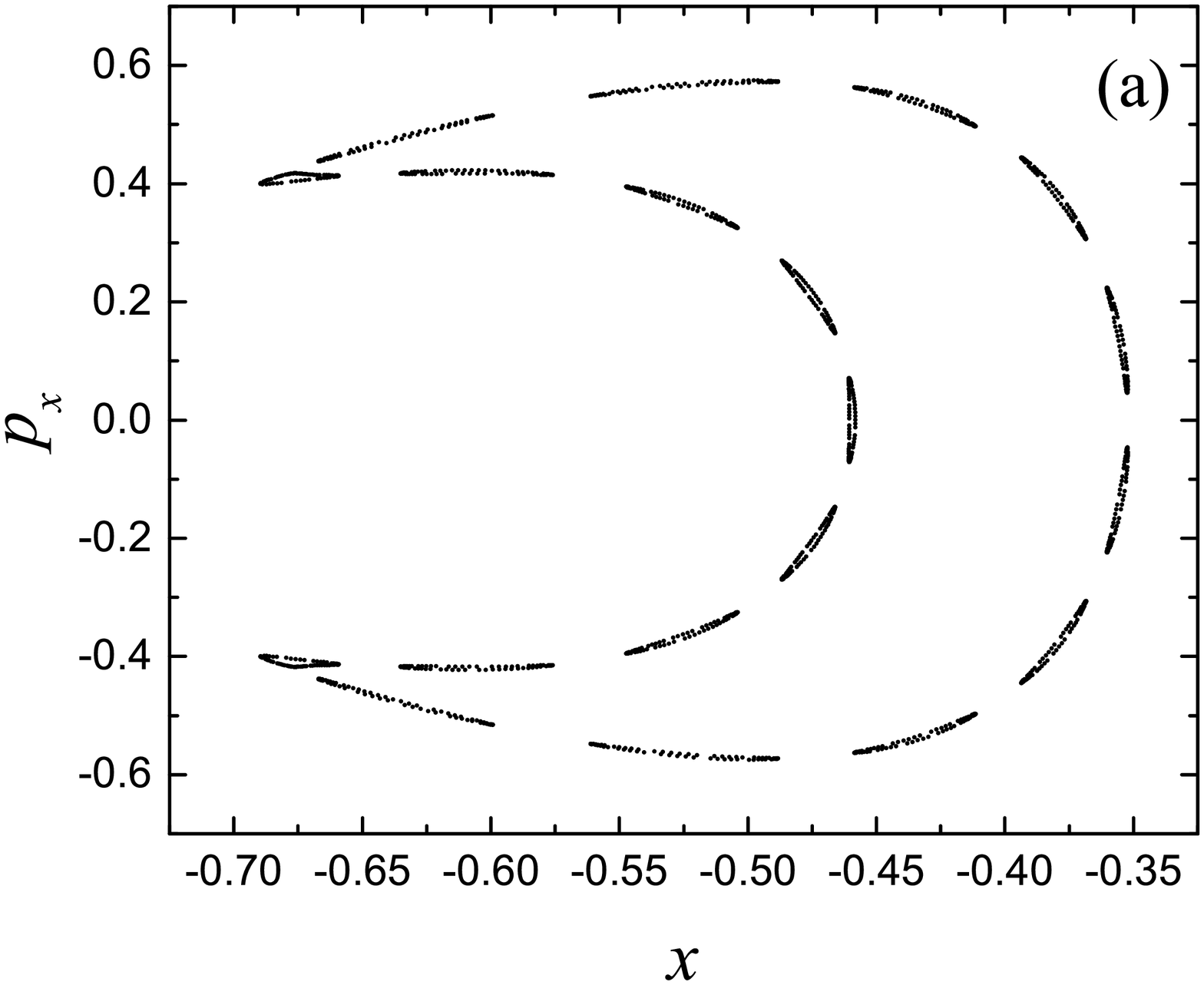}
\includegraphics[scale=0.25]{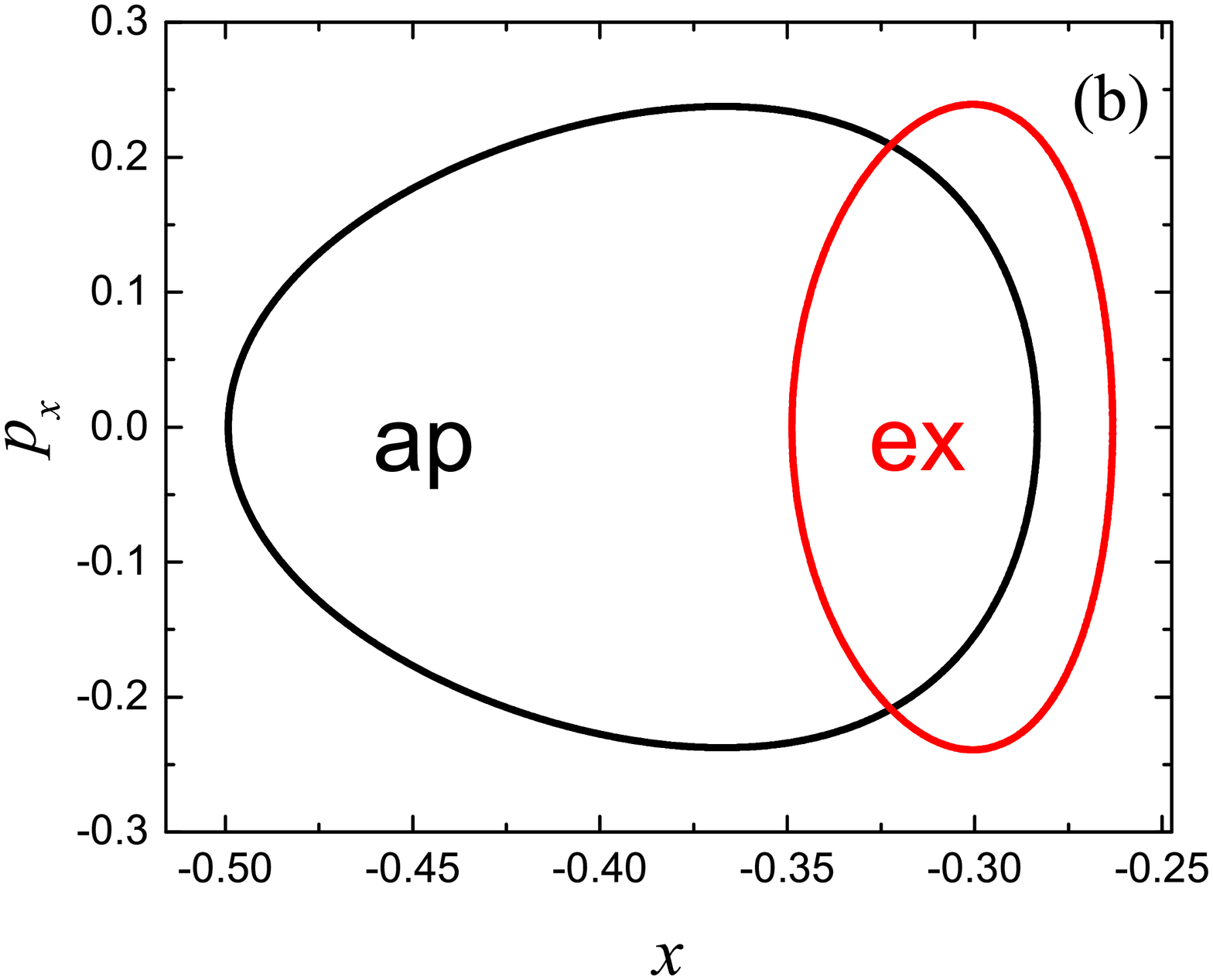}
\includegraphics[scale=0.25]{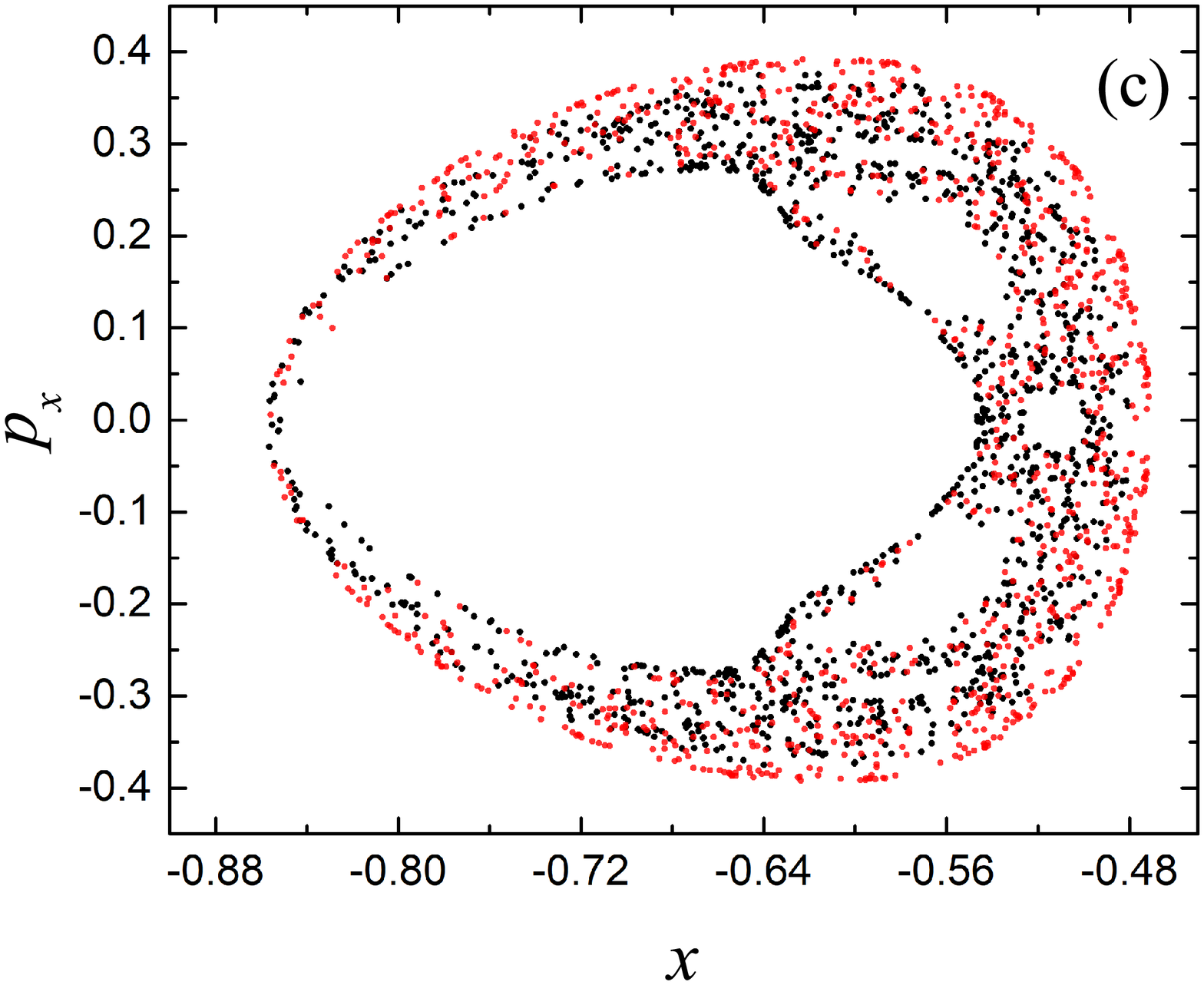}
\includegraphics[scale=0.25]{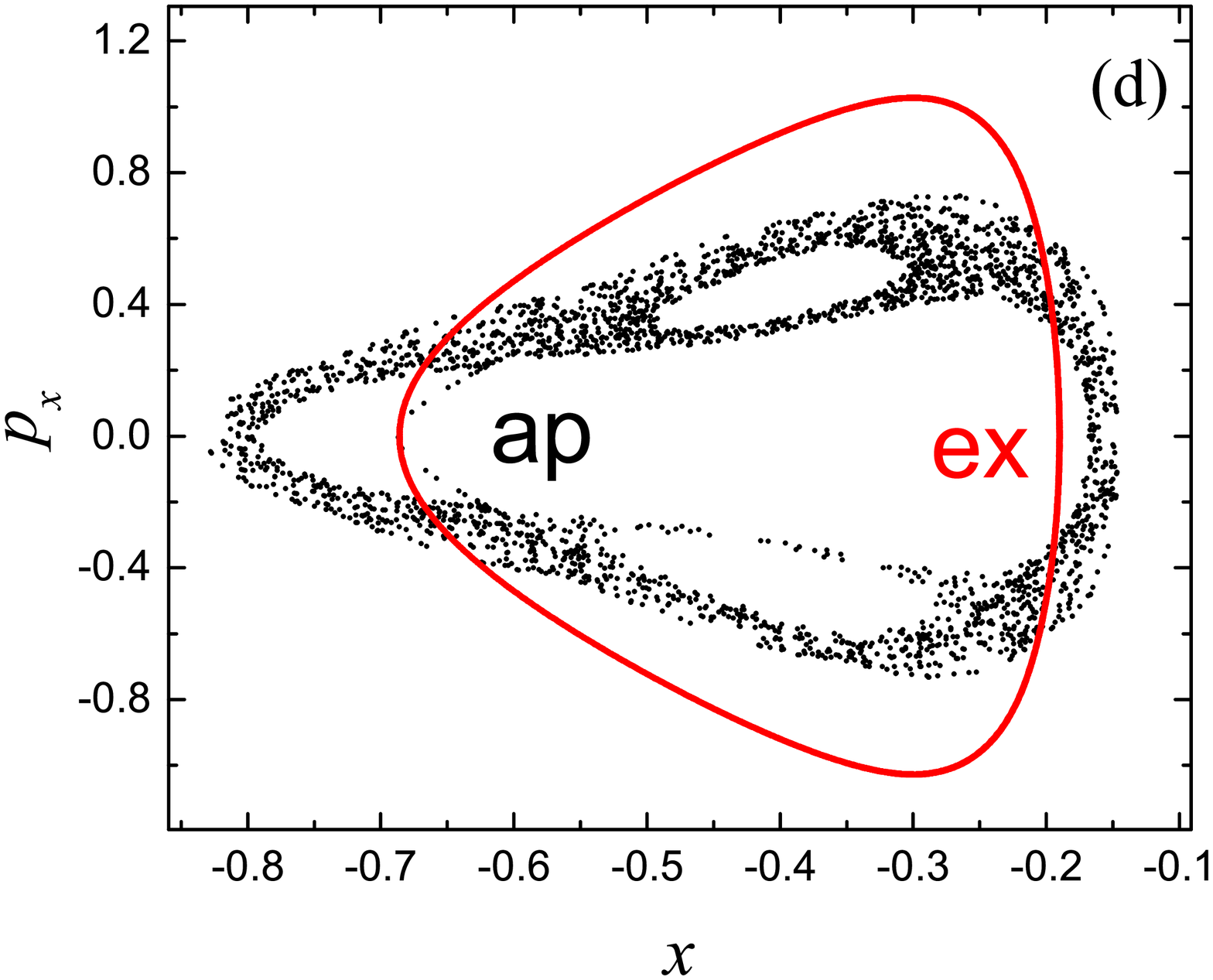}
\caption{ Phase-space structures on the Poincar\'{e} section $y=0$
with $v_y>0$, which are described by RKF89 solving the PN circular
restricted three-body problem. ap represents the approximate
Euler-Lagrange equations, and ex stands for the exact
Euler-Lagrange equations. The parameters are $\mu_{2}=0.001$,
$\mu_{1}=1-\mu_{2}$, $C_{j}=3.07$ and $c=100$, and the initial
conditions are $y=v_{x}=0$. The other initial conditions and
parameters are  (a): $a=1$ and $x=0.35$; (b): $a=10$ and $x=0.28
$; (c): $a=1$ and $x=0.4895$; (d): $a=2$ and $x=0.687$. The
approximate equations and the exact ones yield the same regular
torus in panel (a) and same chaotic solutions in panel (c). They
have different regular tori in panel (b). In panel (d), the
approximate equations correspond to chaos, whereas the exact
equations exhibit the regularity. }
 \label{Fig1}}
\end{figure*}

\begin{figure*}[ptb]
\center{
\includegraphics[scale=0.25]{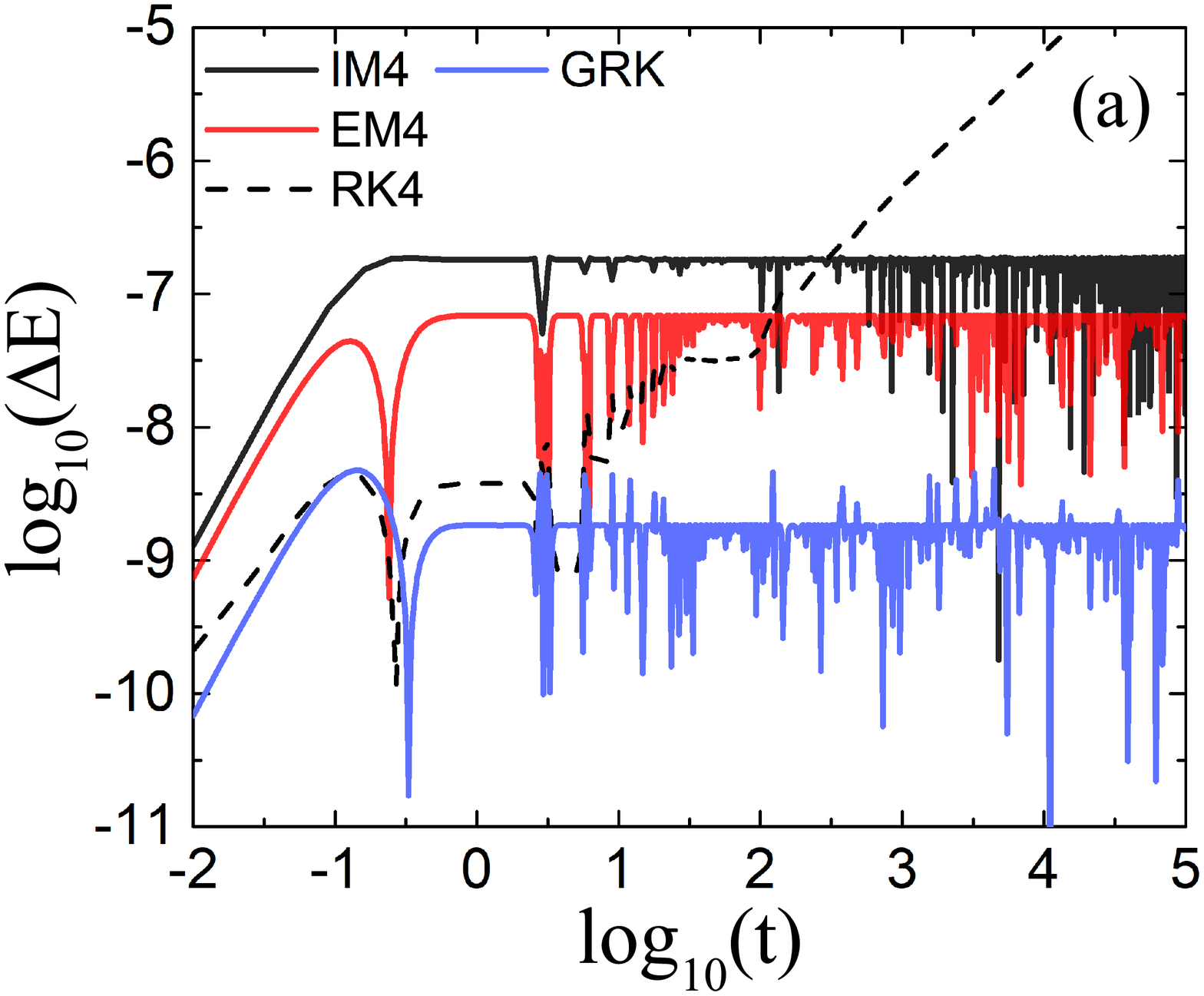}
\includegraphics[scale=0.25]{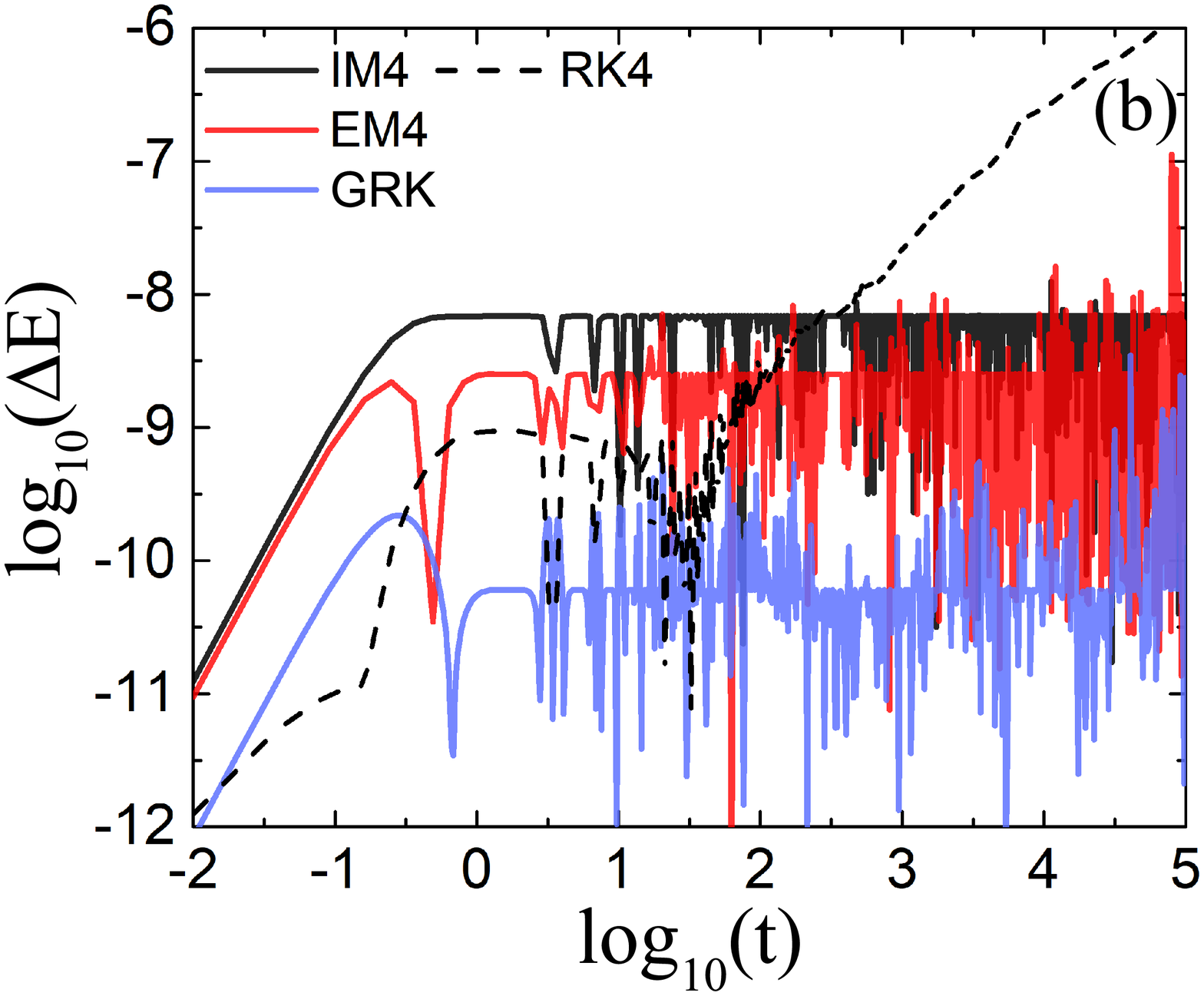}
\includegraphics[scale=0.25]{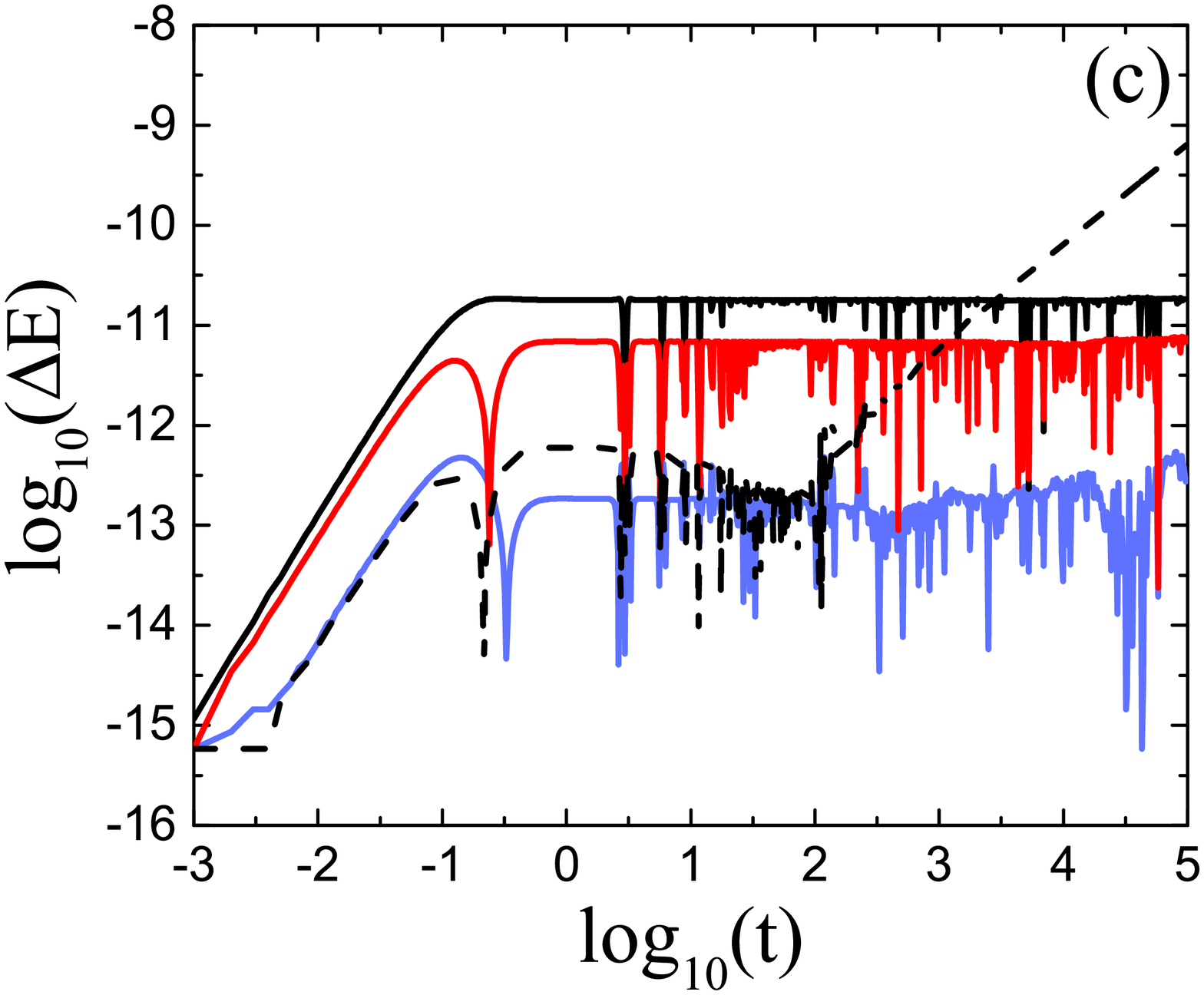}
\includegraphics[scale=0.25]{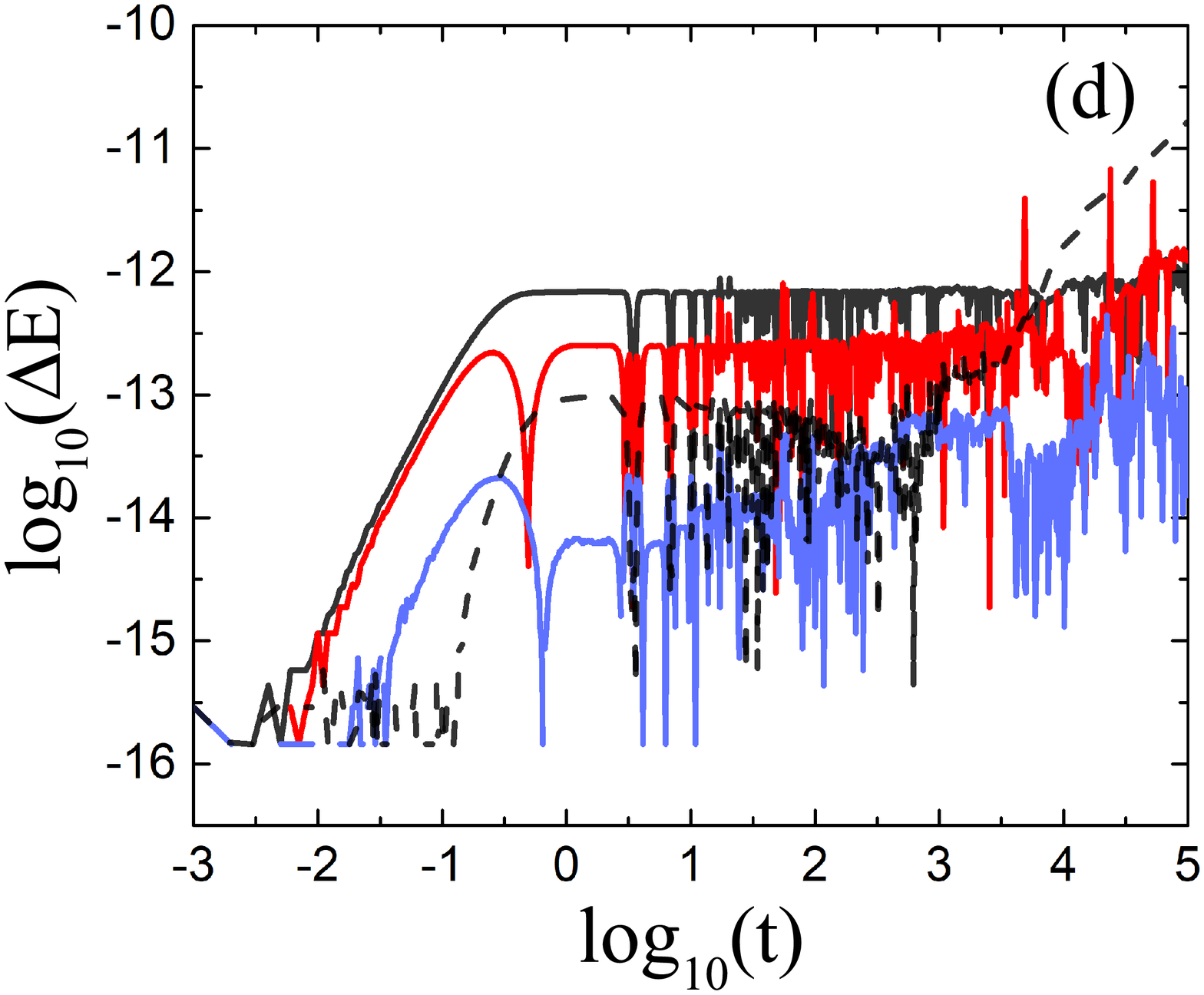}
\caption{{Relative energy errors for four
algorithms solving the exact Euler-Lagrange equations. (a) The
orbit is the regular orbit of Fig. 1(a). (b) The orbit is the
chaotic orbit of Fig. 1(c). The step size is $h=0.01$ in panels
(a) and (b). Panels (c) and (d) that adopt a smaller step size
$h=0.001$ correspond to panels (a) and (b), respectively.} }
 \label{Fig2}}
\end{figure*}

\begin{figure*}[ptb]
\center{
\includegraphics[scale=0.25]{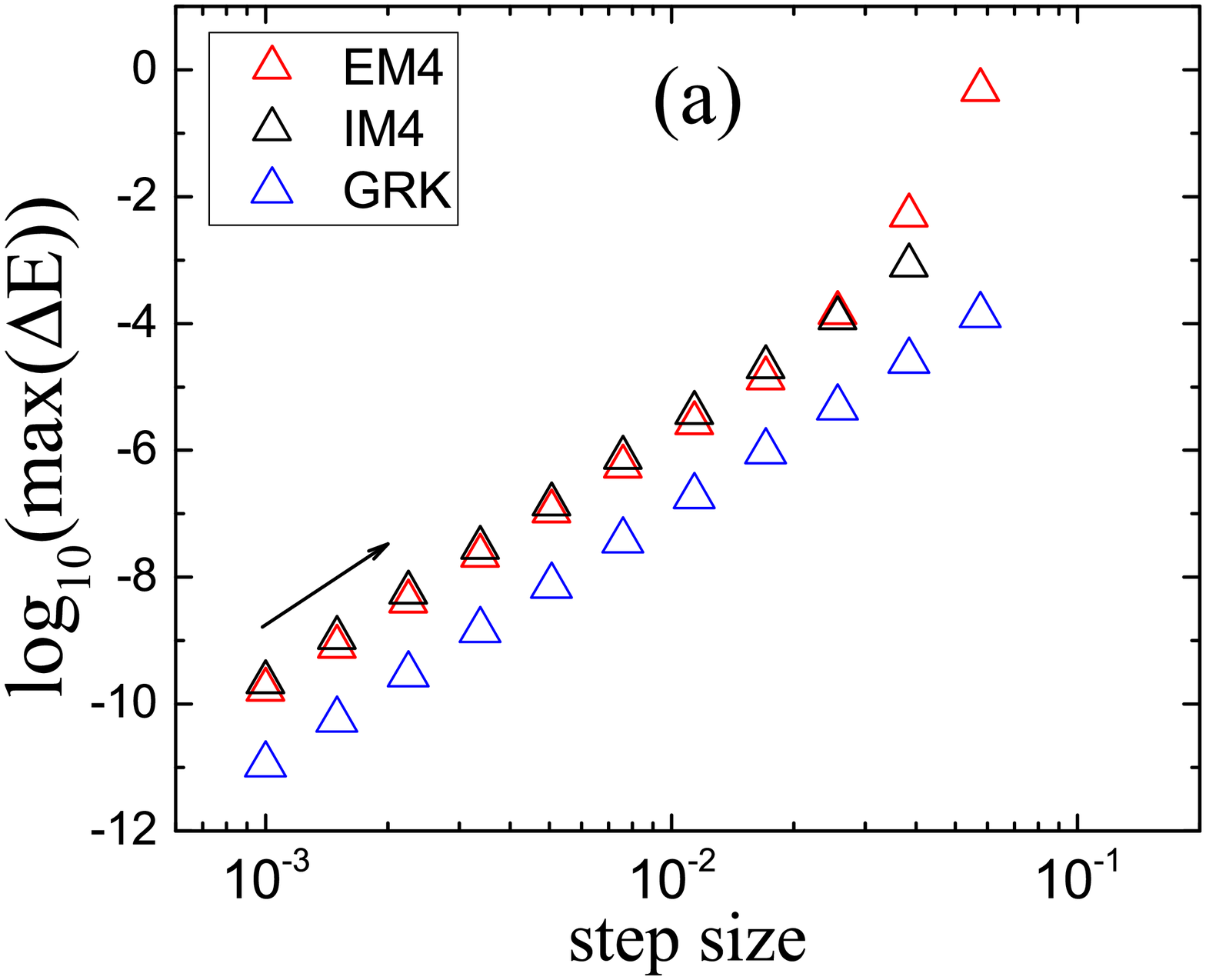}
\includegraphics[scale=0.25]{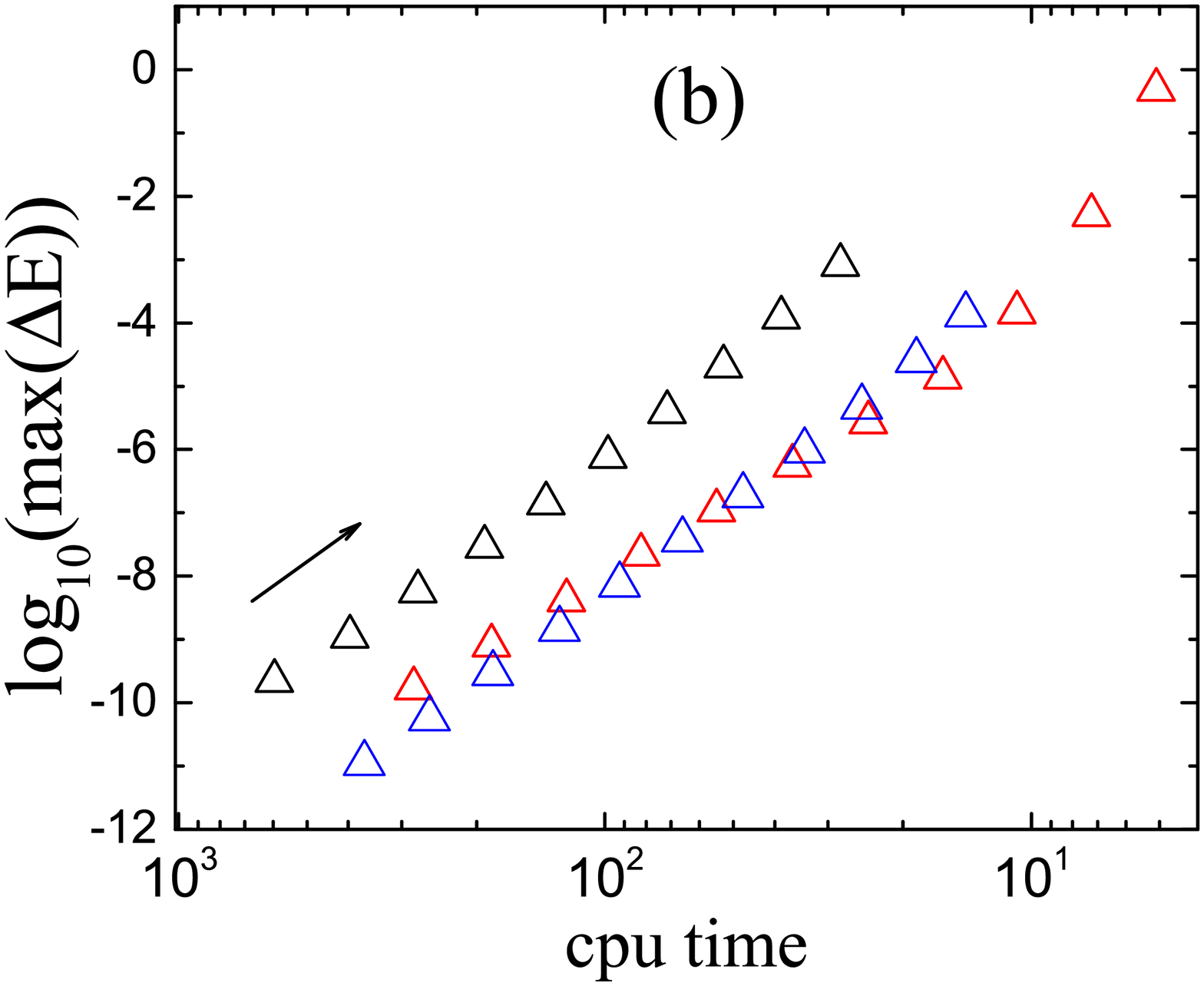}
\caption{ Efficiency plots for the three methods EM4, GRK and IM4.
The initial condition  and parameters are $c=100$, $a=1$ and
$x=0.28$. The maximum relative energy error for each time step and
algorithm is obtained after the integration time $t=10^5$. The
dependence of relative energy error on the step size in (a) shows
that EM4 and IM4 have almost the same accuracies for a given step
size. The dependence of relative energy error on CPU time (unit:
second) in (b) indicates that EM4 takes less computational cost
than IM4 and is almost the same computational cost as GRK for a
given accuracy. }
 \label{Fig3}}
\end{figure*}

\begin{figure*}[ptb]
\center{
\includegraphics[scale=0.3]{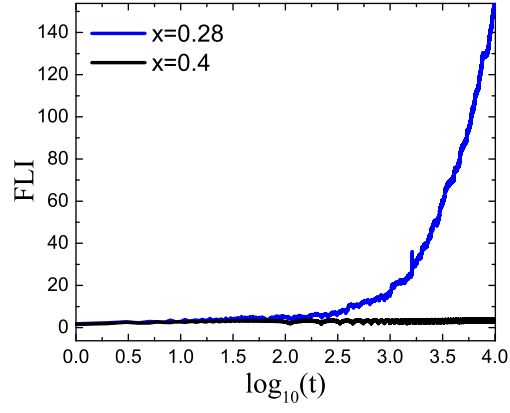}
\caption{ Fast Lyapunov indicators (FLIs) of two orbits. The
parameters are $c=100$, $a=1$ and $c_j=3.07$. }
 \label{Fig4}}
\end{figure*}

\begin{figure*}[ptb]
\center{
\includegraphics[scale=0.25]{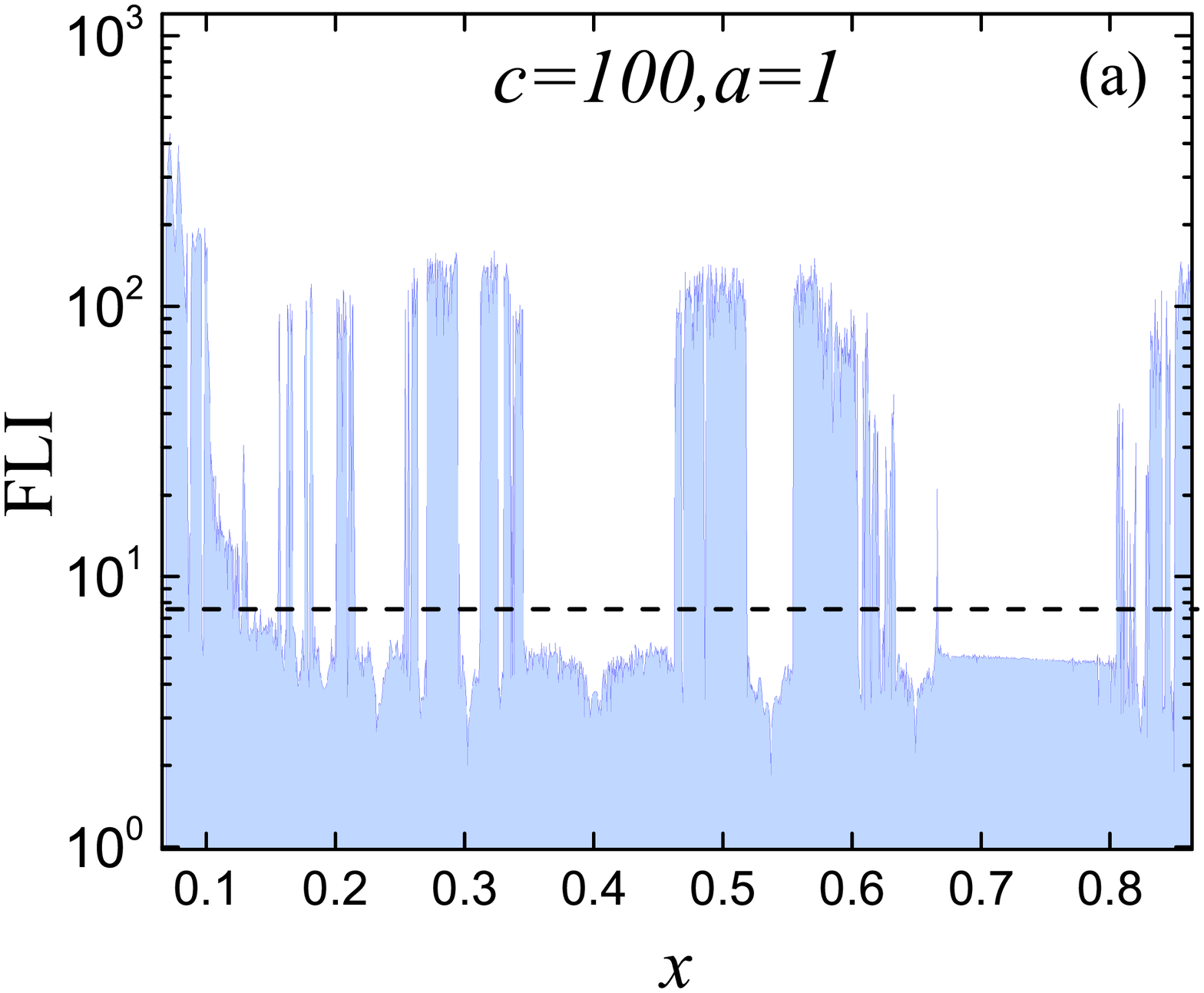}
\includegraphics[scale=0.25]{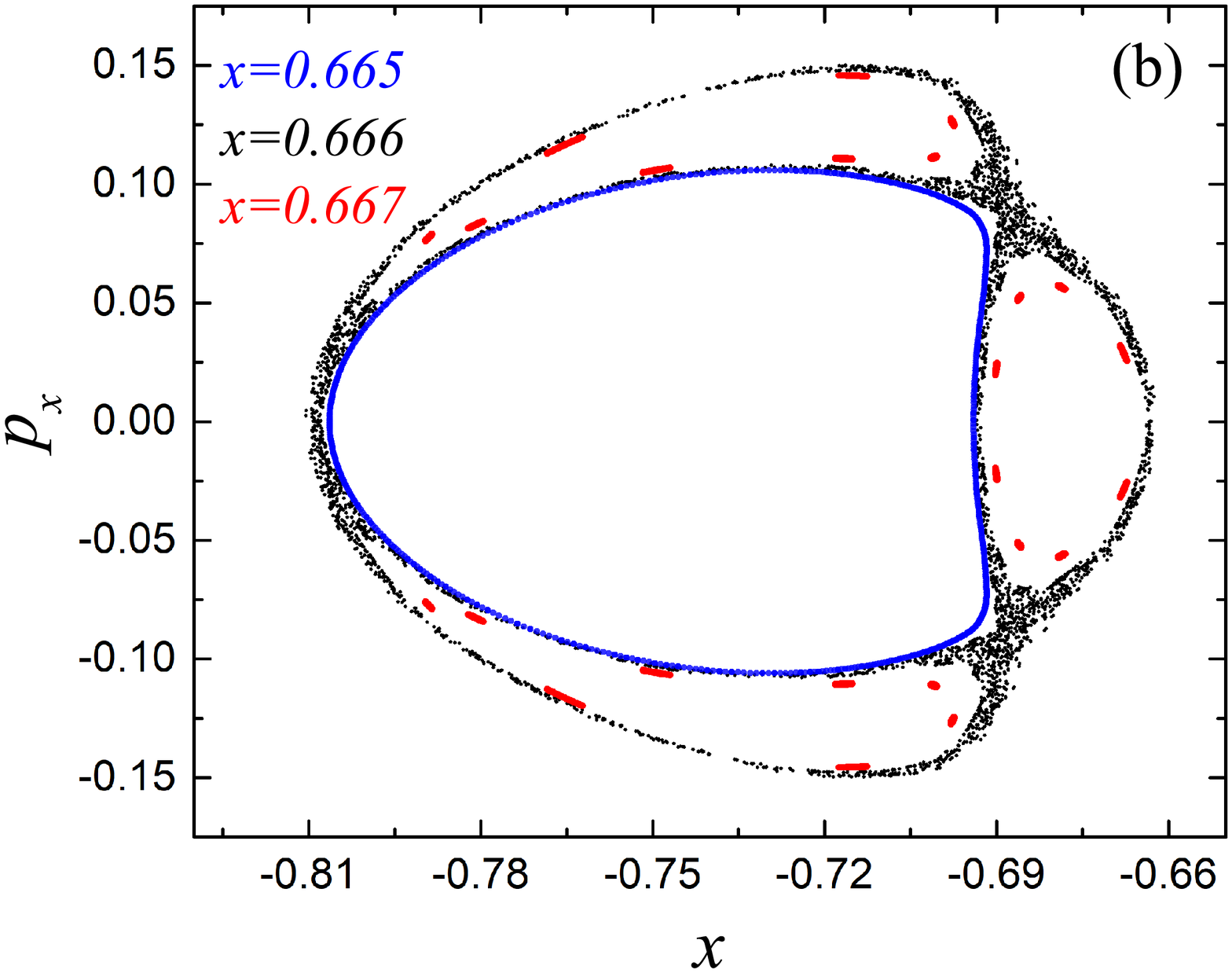}
\caption{ (a) Dependence of FLIs on the initial values $x$. (b)
Poincar\'{e} sections  for ordered orbits with $x=0.665$
corresponding to FLI=5.3 and  $x=0.667$ (FLI=5.2), and a chaotic
orbit with $x =0.666$ (FLI=21.1). }
 \label{Fig5}}
\end{figure*}

\begin{figure*}[ptb]
\center{
\includegraphics[scale=0.25]{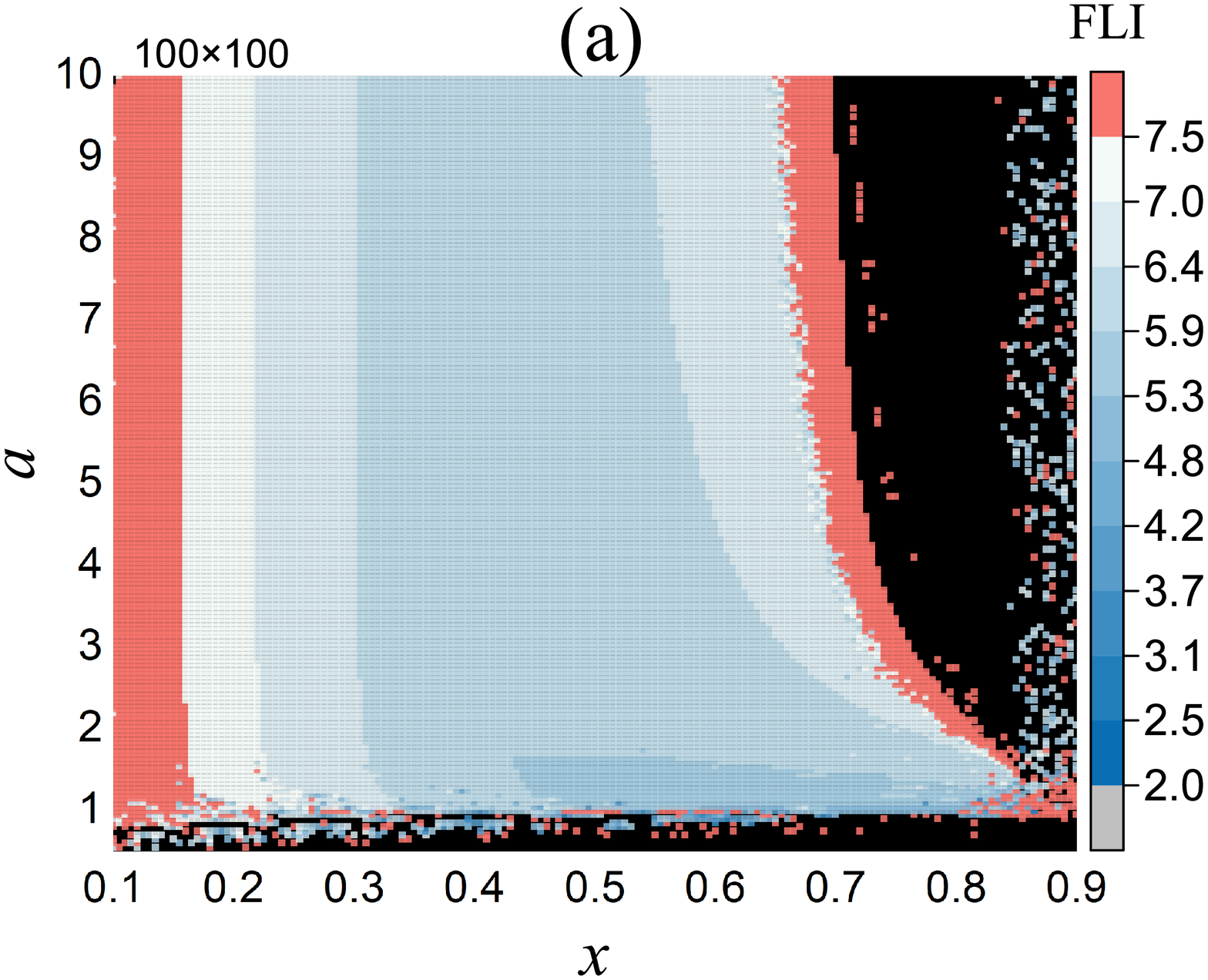}
\includegraphics[scale=0.25]{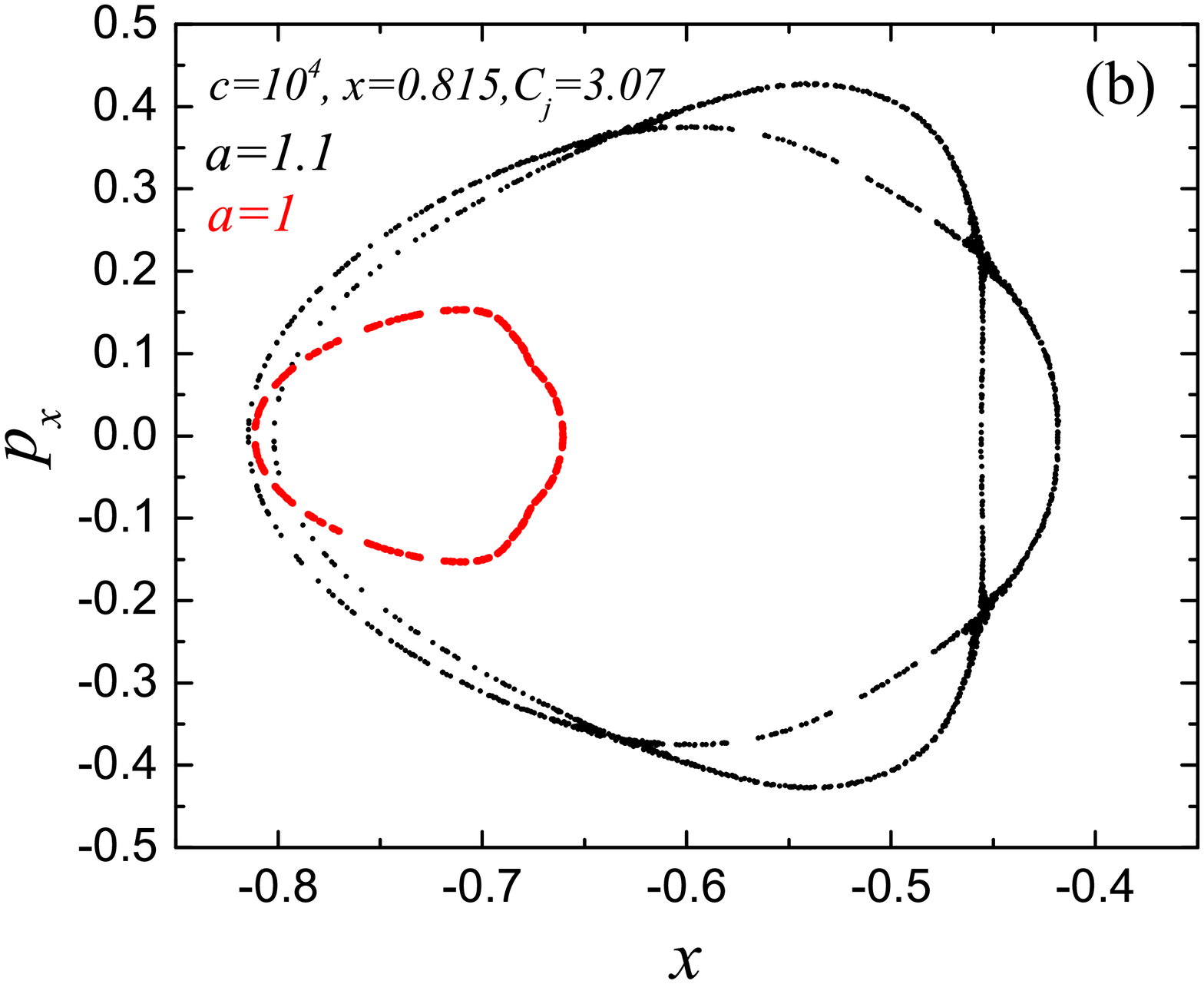}
\includegraphics[scale=0.25]{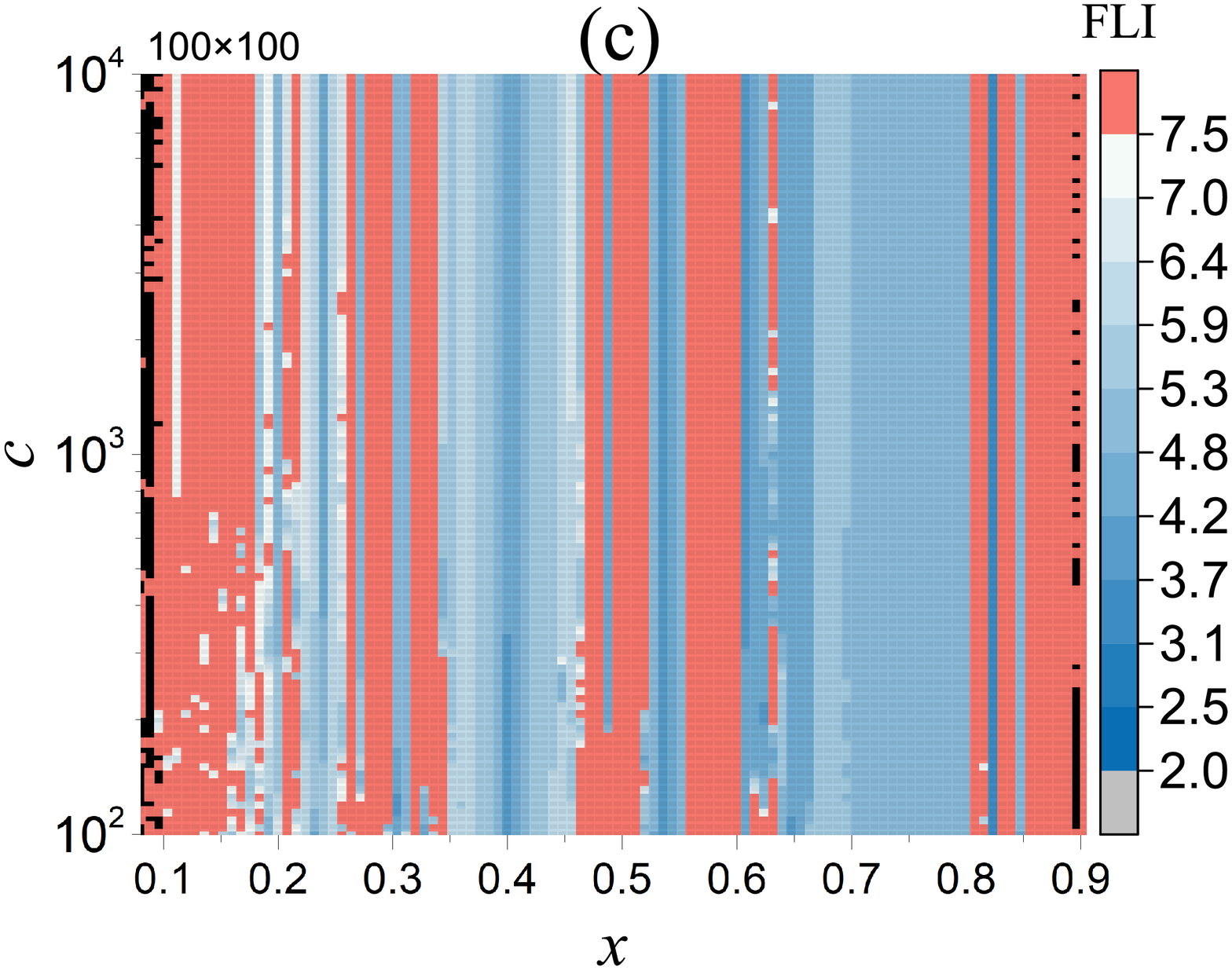}
\includegraphics[scale=0.25]{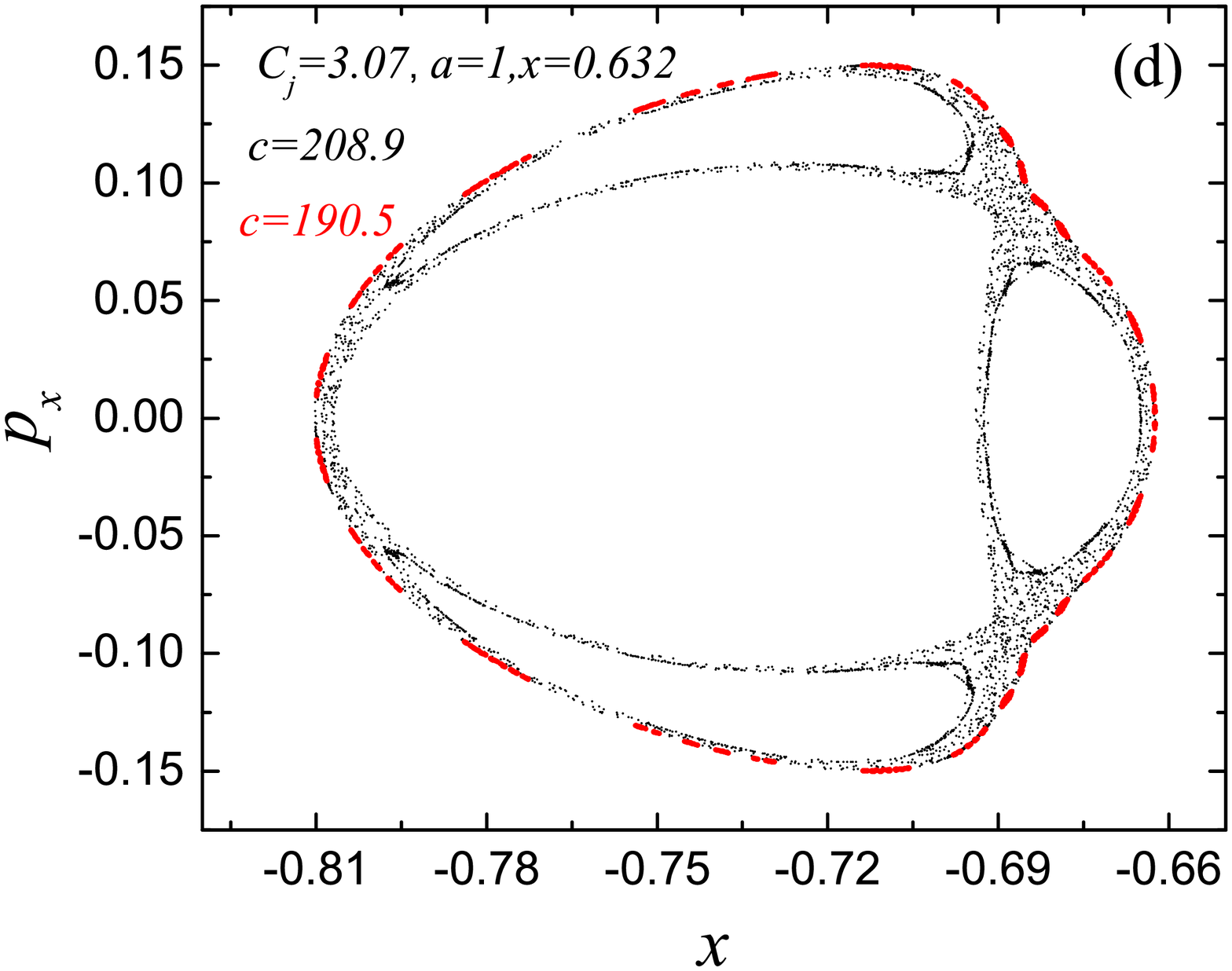}
\caption{(a) Dependence of FLIs on the parameter $a$ and initial
value $x$. $c=10^4$ is fixed. Red: chaos, Black: unstable, other:
order. (b) Poincar\'{e} sections for the regular case of $x$=0.815
and $a=1$ with FLI=5.59 and the chaotic case of $x$=0.815 and
$a=1.1$ with FLI=46.88. (c) Dependence of FLIs on the parameter
$c$ and initial value $x$. $a=1$ is fixed. (d) Poincar\'{e}
sections for the regular case of $x$=0.632 and $c=190.5$ with
FLI=5.91 and the chaotic case of $x$=0.632 and $c=208.9$ with
FLI=12.48.  }
 \label{Fig6}}
\end{figure*}

\begin{figure*}[ptb]
\center{
\includegraphics[scale=0.25]{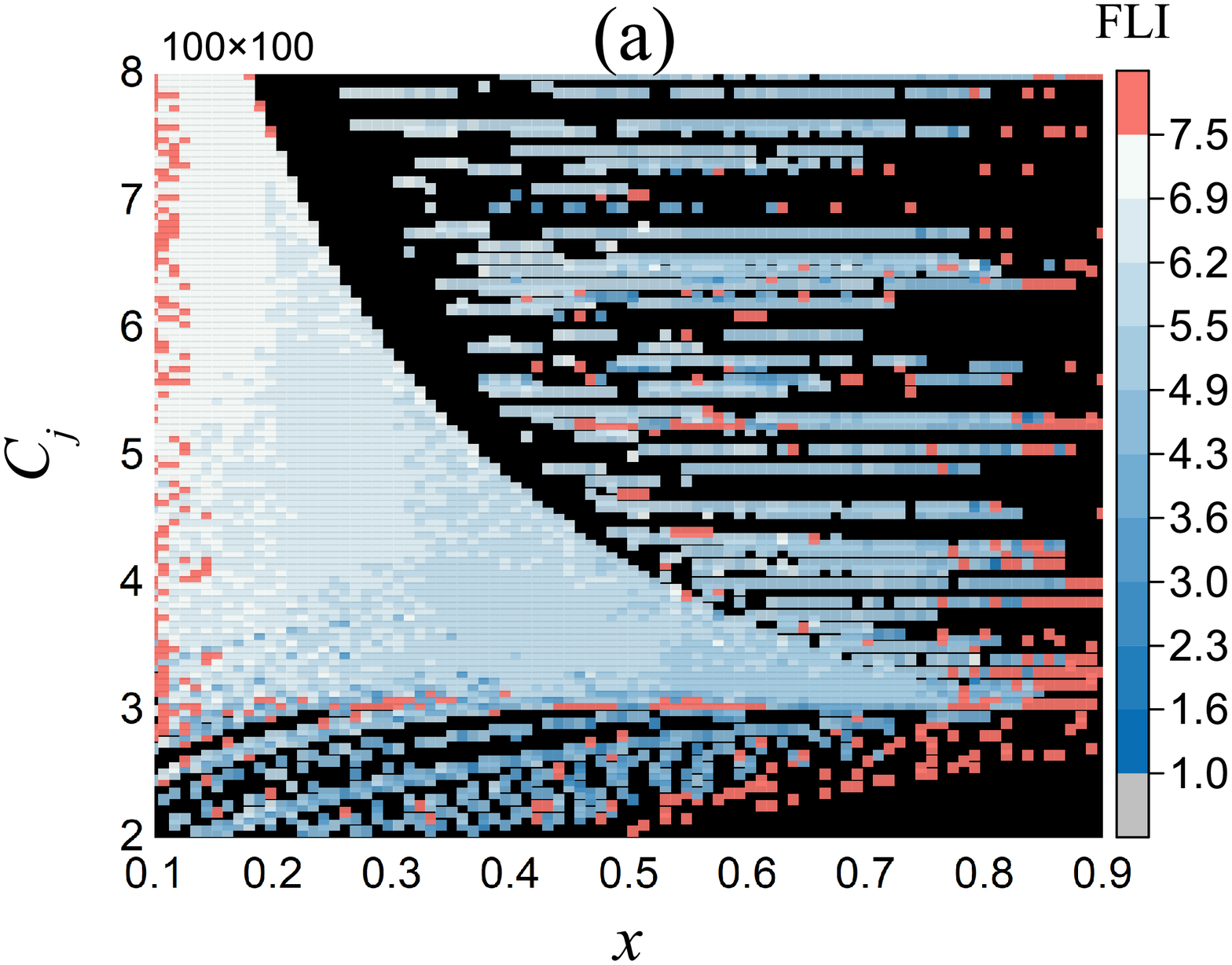}
\includegraphics[scale=0.25]{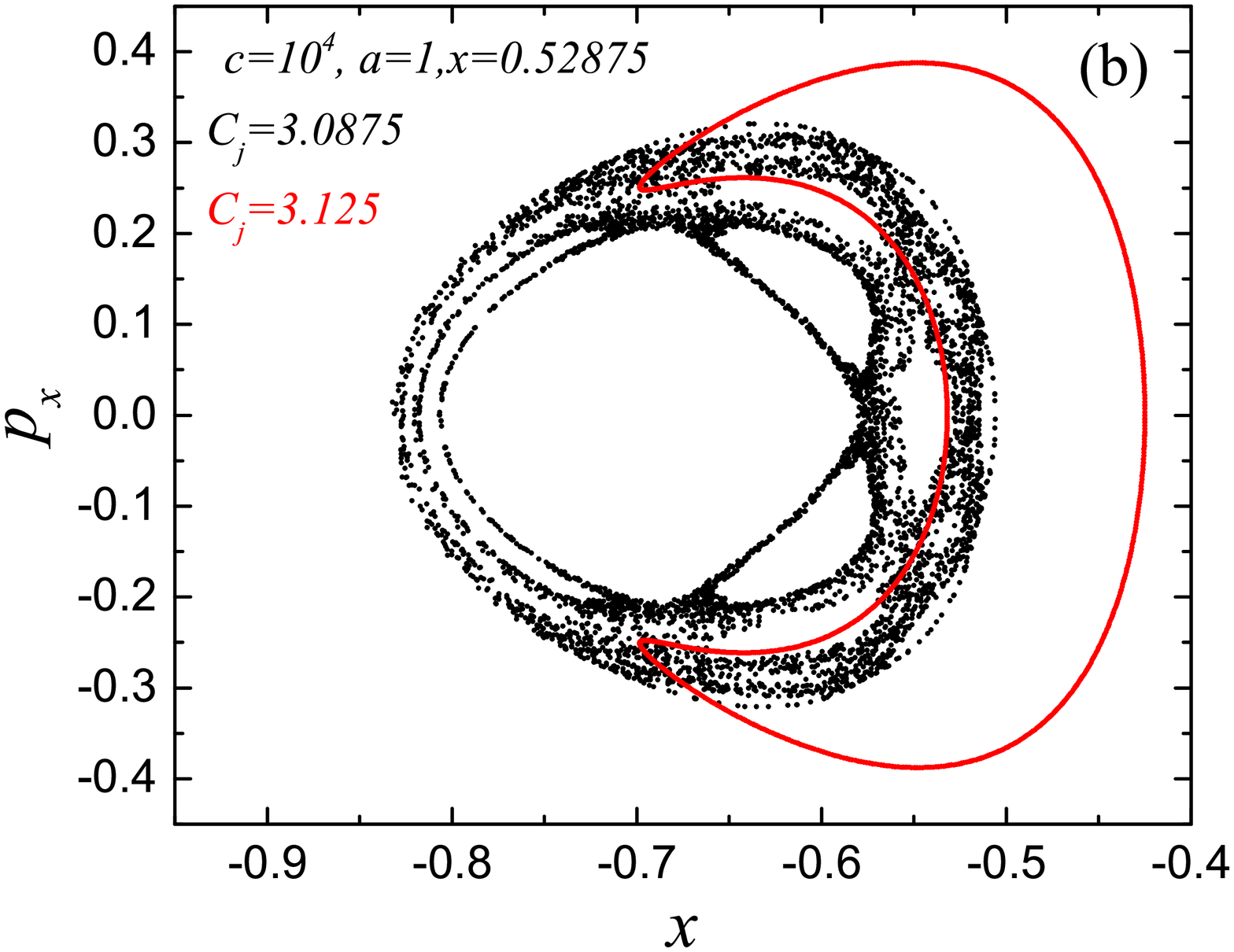}
\caption{ (a) Dependence of FLIs on the parameter $C_j$ and
initial value $x$. (b) Poincar\'{e} sections for two orbits. }
 \label{Fig7}}
\end{figure*}

\begin{figure*}[ptb]
\center{
\includegraphics[scale=0.3]{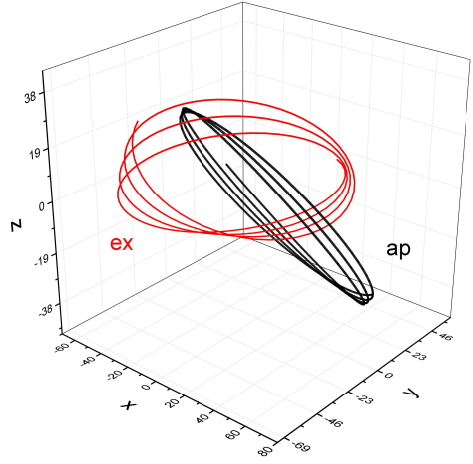}
\caption{ Three-dimensional orbits, described by RKF89 solving the
approximate Euler-Lagrange equations and the exact ones of
spinning compact binaries. }
 \label{Fig8}}
\end{figure*}

\begin{figure*}[ptb]
\center{
\includegraphics[scale=0.20]{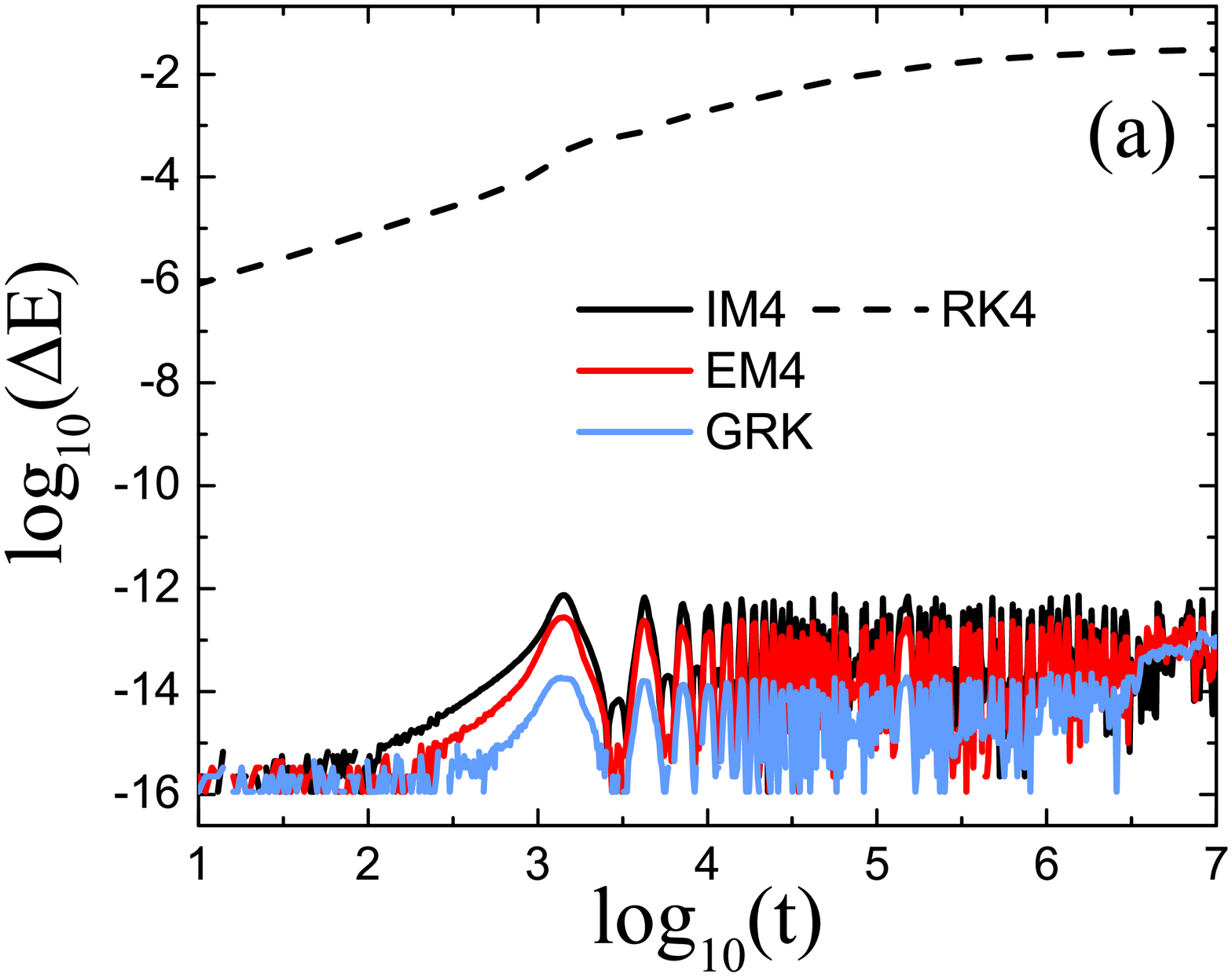}
\includegraphics[scale=0.20]{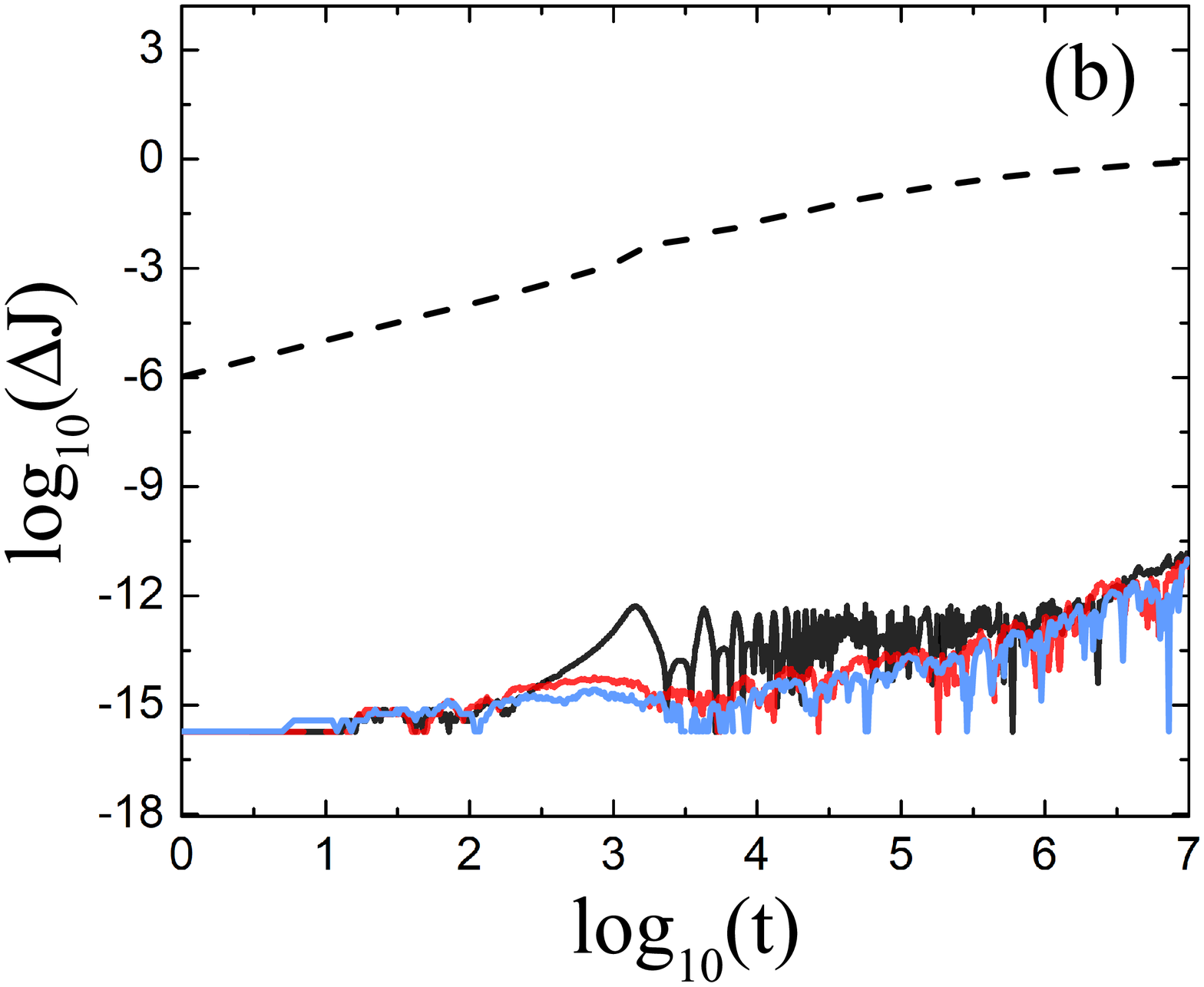}
\includegraphics[scale=0.20]{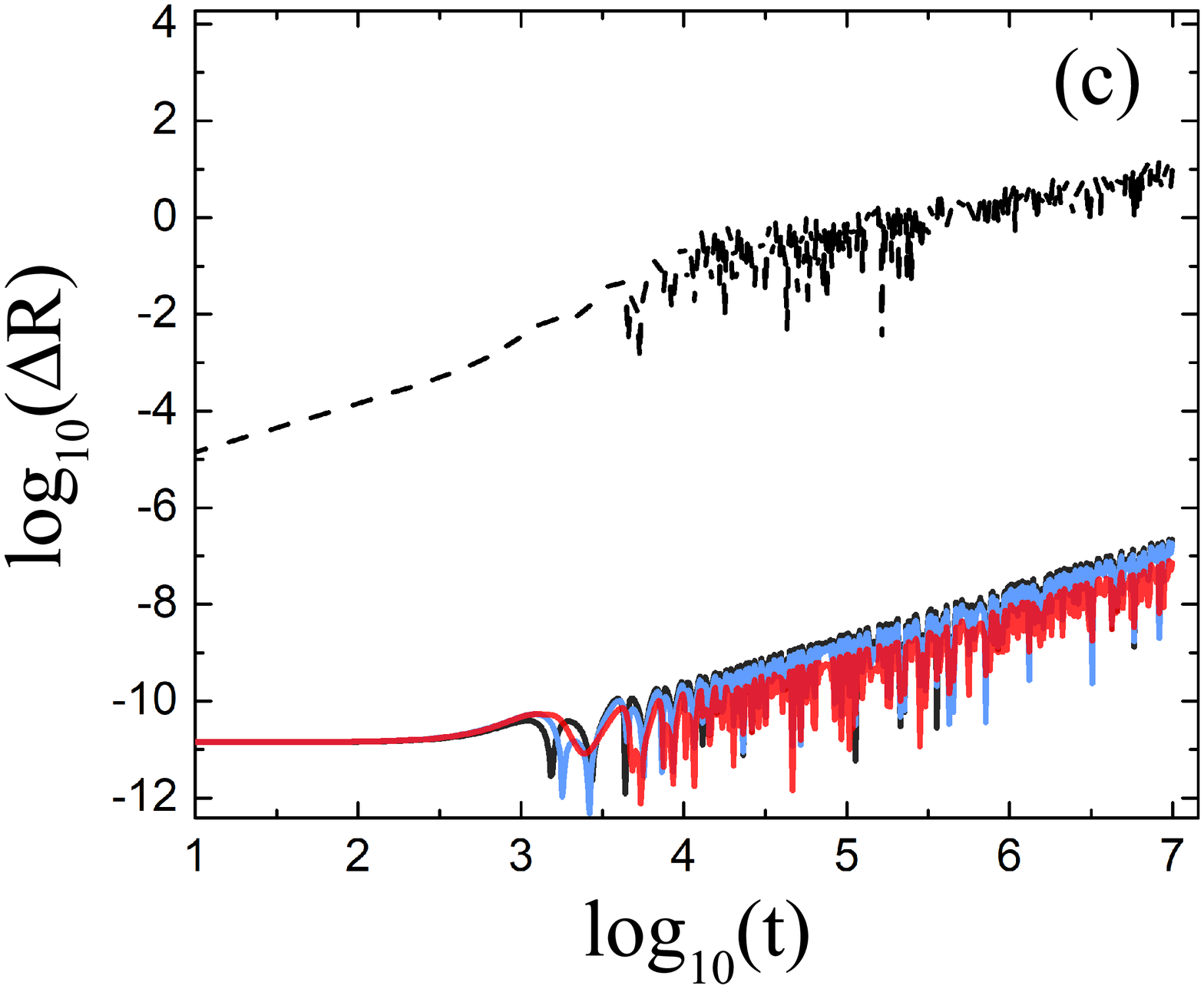}
\caption{ Relative errors in the energy (a), angular momentum (b)
and position (c) for the four algorithms acting on the coherent PN
Euler-Lagrange equations. The initial conditions are $x=70$,
$p_{y}=\sqrt{(1-e)/x}$ with $e=0.16$,
$\theta_{1}=\theta_{2}=\pi/2$, $\xi_{1}=0.1$ and $\xi_{2}=0.95$.
The parameters are $\beta =4$ and $\chi_1=\chi_2=1$. }
 \label{Fig9}}
\end{figure*}

\begin{figure*}[ptb]
\center{
\includegraphics[scale=0.25]{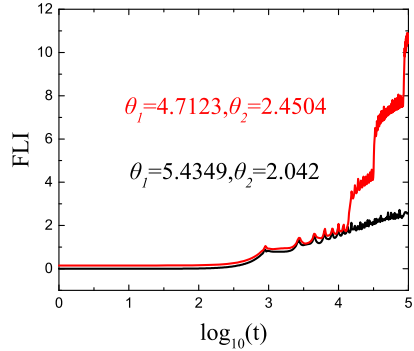}
\caption{ FLIs of two orbits in the exact equations. The initial
conditions are the same as those in Fig. 9, but the initial
eccentricity $e=0.66$ and initial spin angles $\theta_{1}$ and
$\theta_{2}$ are different. For the initial spin angles
$\theta_{1}=5.4349$ and $\theta_{2}=2.042$ with FLI= 2.56, the
dynamics is regular. However, the dynamics is chaotic for the
initial spin angles $\theta_{1}=4.7123$ and $\theta_{2}=2.4504$
with FLI= 10.48. }
 \label{Fig10}}
\end{figure*}

\begin{figure*}[ptb]
\center{
\includegraphics[scale=0.30]{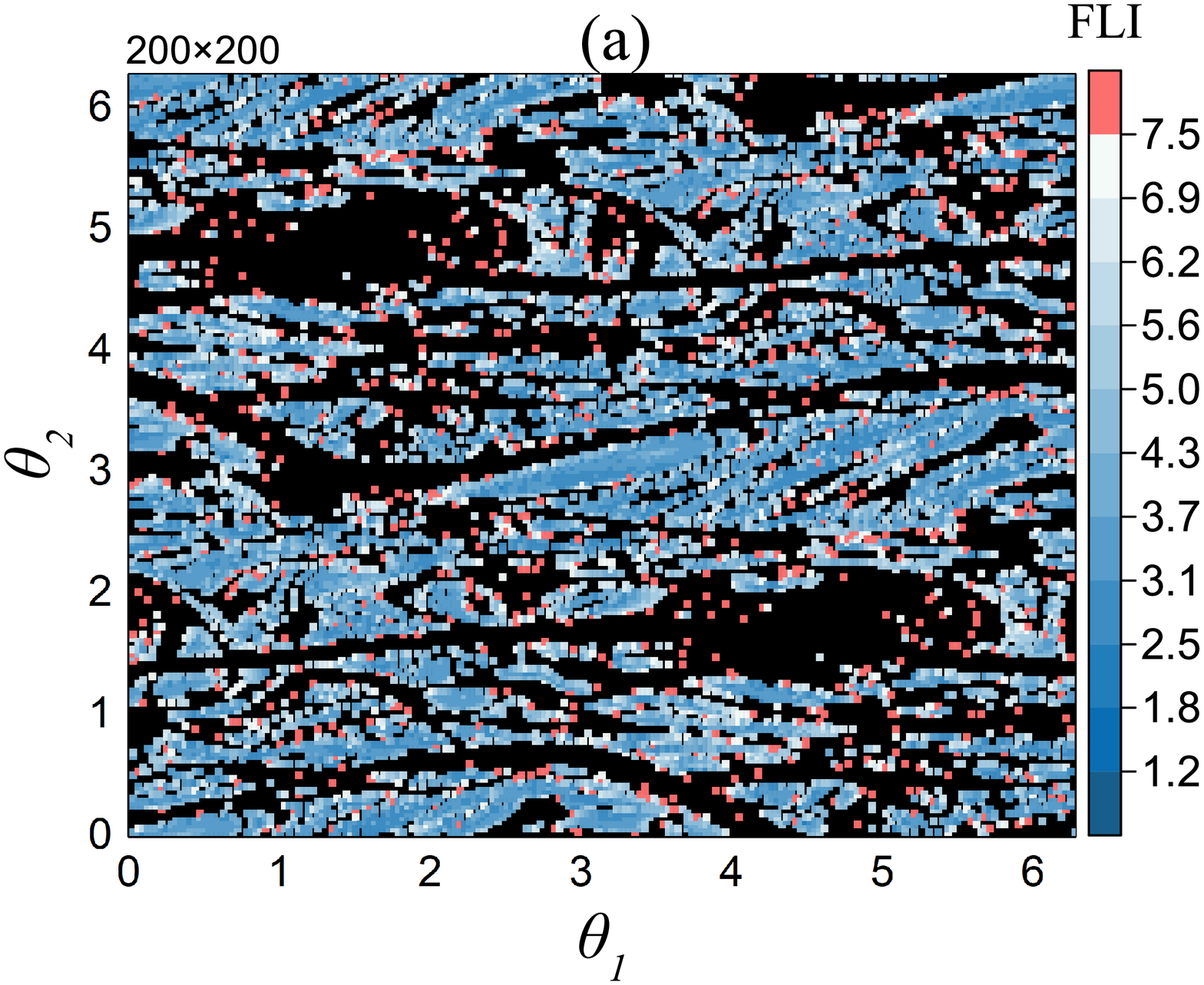}
\includegraphics[scale=0.30]{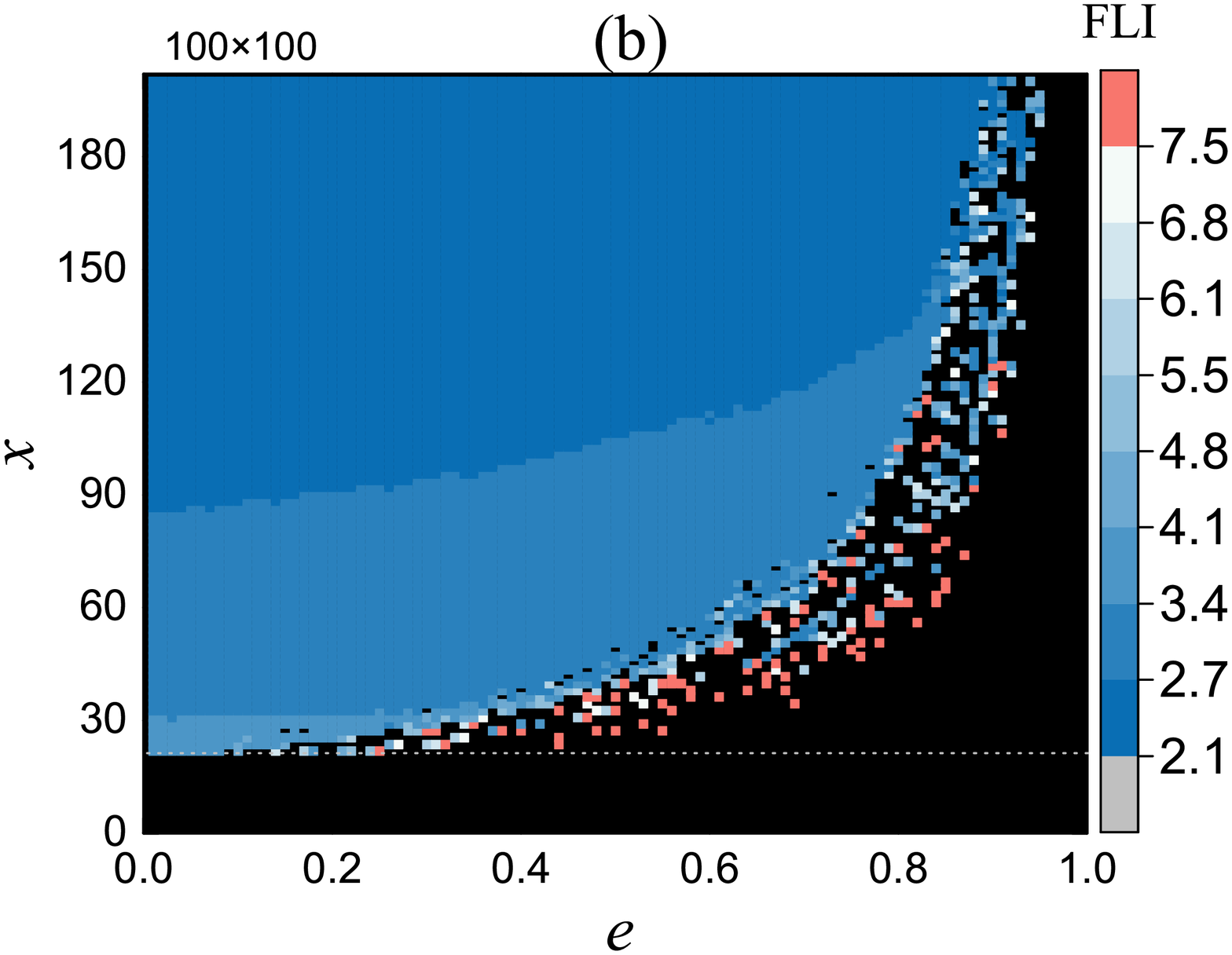}
\caption{ (a) Dynamical structures by using the FLIs to scan the
initial spin angles $\theta_{1}$ and $\theta_{2}$. (b) To scan the
initial eccentricities $e$ and initial separations $x$. Red:
chaos, Black: unstable, Blue: order. }
 \label{Fig11}}
\end{figure*}

\end{document}